\pdfoutput=1
\documentclass[prd,showpacs,letterpaper,10pt,aps,notitlepage,nofootinbib,superscriptaddress,preprintnumbers]{revtex4-1}

\usepackage{amsfonts,amsmath,graphicx}
\usepackage{hyperref}
\usepackage{color}
\usepackage{xcolor}
\usepackage{placeins}
\usepackage{diagbox}
\usepackage{multirow}
\usepackage[utf8]{inputenc}
\usepackage[T1]{fontenc}

\hypersetup{colorlinks, linkcolor = [rgb]{0,0.0,0.75}, citecolor = [rgb]{0,0.0,0.75}, urlcolor = [rgb]{0,0.0,0.75}}

\newcommand{\C}{\mathcal{C}}

\newcommand{\xb}{\textbf{x}}
\newcommand{\yb}{\textbf{y}}

\newcommand{\op}{O}

\newcolumntype{C}[1]{>{\centering\arraybackslash}m{#1}}

\definecolor{darkgreen}{rgb}{0,0.5,0}

\newcommand{\bss}[1]{\ensuremath{{\boldsymbol{#1}}}}

\begin{document}

\preprint{JLAB-THY-19-2912}
\preprint{RBRC-1307}

\title{\texorpdfstring{Lattice QCD investigation of a doubly-bottom $\bar{b} \bar{b} u d$ tetraquark \\ with quantum numbers $I(J^P) = 0(1^+)$}{Lattice QCD investigation of a doubly-bottom \bar{b} \bar{b} u d tetraquark with quantum numbers I(J P) = 0(1 +)}}

\author{Luka Leskovec}
\affiliation{Thomas Jefferson National Accelerator Facility, 12000 Jefferson Avenue, Newport News, Virginia 23606, USA}

\author{Stefan Meinel}
\affiliation{Department of Physics, University of Arizona, Tucson, Arizona 85721, USA}
\affiliation{RIKEN BNL Research Center, Brookhaven National Laboratory, Upton, New York 11973, USA}

\author{Martin Pflaumer}
\affiliation{Goethe-Universit\"at Frankfurt am Main, Institut f\"ur Theoretische Physik, Max-von-Laue-Stra{\ss}e 1, D-60438 Frankfurt am Main, Germany}

\author{Marc Wagner}
\affiliation{Goethe-Universit\"at Frankfurt am Main, Institut f\"ur Theoretische Physik, Max-von-Laue-Stra{\ss}e 1, D-60438 Frankfurt am Main, Germany}

\date{June 3, 2019}

\begin{abstract}

We use lattice QCD to investigate the spectrum of the $\bar{b} \bar{b} u d$ four-quark system with quantum numbers $I(J^P) = 0(1^+)$. We use five different gauge-link ensembles with $2+1$ flavors of domain-wall fermions, including one at the physical pion mass, and treat the heavy $\bar{b}$ quark within the framework of lattice nonrelativistic QCD. Our work improves upon previous similar computations by considering in addition to local four-quark interpolators also nonlocal two-meson interpolators and by performing a L\"uscher analysis to extrapolate our results to infinite volume. We obtain a binding energy of $(-128 \pm 24 \pm 10) \, \textrm{MeV}$, corresponding to the mass $(10476 \pm 24 \pm 10) \, \textrm{MeV}$, which confirms the existence of a $\bar{b} \bar{b} u d$ tetraquark that is stable with respect to the strong and electromagnetic interactions.
\end{abstract}

\maketitle

% ********************
% ********************
% ********************
% ********************
% ********************

\section{Introduction}

Mesons, i.e., hadrons with integer spin, were first envisioned by Gell-Mann and Zweig \cite{GellMann:1964nj,Zweig:1964jf} to be built from one, two or more quark-antiquark pairs. However, systems that manifestly contain more than a single quark-antiquark pair were found only relatively recently, primarily in the heavy-quark sector \cite{Belle:2011aa, Olsen:2015zcy,Lebed:2016hpi,Esposito:2016noz,Richard:2016eis,Olsen:2017bmm}. Exotic mesons can be characterized as having $J^{PC}$ quantum numbers
that cannot be constructed in the simple quark-antiquark model, or as having a manifestly exotic quark flavor content. In this work, we consider
an example for the latter, a $\bar{b} \bar{b} u d$ tetraquark.\footnote{In the literature, the term ``tetraquark'' is somewhat ambiguous. In certain papers it exclusively refers to a diquark-antidiquark structure, while in other papers it is used more generally for arbitrary bound states and resonances with a strong four-quark component, including, e.g., mesonic molecules. Throughout this paper we follow the latter convention. Moreover, the $\bar{b} \bar{b} u d$ system is a tetraquark in a fully rigorous sense, since it contains four net quark flavors.}

It can be shown that QCD-stable $\bar{Q}\bar{Q}qq$ tetraquarks must exist in the limit $m_Q\to \infty$ \cite{Carlson:1987hh, Manohar:1992nd, Eichten:2017ffp}. In this limit, the two heavy antiquarks form a color-triplet object with a size of order $(\alpha_s m_Q)^{-1}$ and a binding energy of order $\alpha_s^2 m_Q$ due to the attractive Coulomb potential at short distances. The doubly-heavy $\bar{Q}\bar{Q}qq$ tetraquarks then become related to singly-heavy $Qqq$ baryons, just like doubly-heavy $\bar{Q}\bar{Q}\bar{q}$ baryons become related to singly-heavy $Q\bar{q}$ mesons \cite{Savage:1990di, Brambilla:2005yk, Cohen:2006jg, Mehen:2017nrh}. The question is whether the physical bottom quark is heavy enough for $\bar{b} \bar{b} qq $ bound states to exist below the $\bar{b}q $-$\bar{b} q$ two-meson thresholds. Studies based on potential models, effective field theories, and QCD sum rules suggest that this is indeed the case \cite{Carlson:1987hh, Manohar:1992nd, SilvestreBrac:1993ss, Brink:1998as, Vijande:2003ki, Janc:2004qn, Vijande:2006jf, Navarra:2007yw, Ebert:2007rn, Zhang:2007mu, Lee:2009rt, Karliner:2017qjm, Eichten:2017ffp, Wang:2017uld, Richard:2018yrm, Park:2018wjk, Wang:2018atz, Liu:2019stu}. Possible experimental search strategies for bottomness-$2$ tetraquarks are discussed in Refs.~\cite{Moinester:1995fk, Ali:2018ifm, Ali:2018xfq}.

Within lattice QCD, $\bar{b} \bar{b} q q$ four-quark systems were explored for the first time using static $\bar{b}$ quarks and the Born-Oppenheimer approximation. A stable $\bar{b} \bar{b} u d$ tetraquark with quantum numbers $I(J^P) = 0(1^+)$ around $30 \ldots 90 \, \textrm{MeV}$ below the $B B^\ast$ threshold as well as a $\bar{b} \bar{b} u d$ tetraquark resonance with quantum numbers $I(J^P) = 0(1^-)$ around $15 \, \textrm{MeV}$ above the $B B$ threshold were predicted \cite{Bicudo:2012qt,Brown:2012tm,Bicudo:2015kna,Bicudo:2016ooe,Bicudo:2017szl}. Effects from the heavy-quark spin were investigated for the stable $I(J^P) = 0(1^+)$ tetraquark by solving a coupled-channel Schr\"odinger equation in Ref.~\cite{Bicudo:2016ooe}. Moreover, several flavor combinations were explored and no stable $\bar{b} \bar{b} q q$ tetraquarks with $q q = s s$ and $q q = c c$ were found in this approach \cite{Bicudo:2015vta}. Recently, the same $\bar{b} \bar{b} q q$ four-quark systems have been investigated with $\bar{b}$ quarks of finite mass treated within nonrelativistic QCD (NRQCD). A stable $\bar{b} \bar{b} u d$ tetraquark with quantum numbers $I(J^P) = 0(1^+)$ was also seen in two such computations \cite{Francis:2016hui,Junnarkar:2018twb}, but there is a quantitative difference by a factor $\approx 2 \ldots 3$ in the binding energy between Refs.\ \cite{Francis:2016hui,Junnarkar:2018twb} and Ref.\ \cite{Bicudo:2016ooe}, which is not yet understood. Moreover, $\bar{Q} \bar{Q} q q$ systems with further flavor combinations $\bar{Q} \bar{Q} \in \{ \bar{b} \bar{b},\bar{b} \bar{c},\bar{c} \bar{c} \}$ and $q \in \{ u,d,s,c \}$ have been investigated and some indication has been obtained that systems with $J^P = 1^+$ and $\bar{Q} \bar{Q} q q \in \{ \bar{b} \bar{b} u s , \bar{b} \bar{b} u c , \bar{b} \bar{b} s c \ , \ \bar{b} \bar{c} u d \ , \  \bar{c} \bar{c} u d \}$ are stable as well \cite{Francis:2018jyb,Junnarkar:2018twb}.

In this paper we perform a lattice QCD study of the $\bar{b} \bar{b} u d$ four-quark system with quantum numbers $I(J^P) = 0(1^+)$, using NRQCD $\bar{b}$ quarks and domain-wall light quarks (results obtained at an early stage of this project have been presented in Ref.~\cite{Peters:2016isf}). We make use of both local interpolating fields (in which the four quarks are jointly projected to zero momentum) and nonlocal interpolating fields (in which each of the two quark-antiquark pairs forming a color-singlet is projected to zero momentum individually). It has been shown in previous studies of other systems \cite{Mohler:2013rwa,Lang:2014yfa} that including both types of interpolating fields is required to reliably determine ground-state energies in exotic channels. In this way we expand on the works of Refs.~\cite{Francis:2016hui,Francis:2018jyb,Junnarkar:2018twb}, where nonlocal interpolating fields were not considered. Having both local and nonlocal interpolating fields allows us to determine the ground-state and first-excited-state energy in the $I(J^P) = 0(1^+)$ channel and perform a L\"uscher analysis of $BB^*$ scattering.

The paper is structured as follows. In Sec.~\ref{sec:latticesetup} we summarize our lattice setup, including the computation of quark propagators. In Sec.~\ref{sec:interpolators} we discuss the interpolating operators and the corresponding correlation functions. The extraction of the energy levels on the lattice is discussed in Secs.~\ref{sec:Bmasses} and \ref{sec:bbudEnergies}. In Sec.~\ref{sec:scatteringAnalysis} we present the scattering analysis, and in Sec.~\ref{sec:errors} we perform a fit of the pion-mass dependence of the binding energy and estimate systematic uncertainties. Our conclusions are given in Sec.~\ref{sec:conclusions}.

% ********************
% ********************
% ********************
% ********************
% ********************

\section{\label{sec:latticesetup}Lattice Setup}

% ********************

\subsection{\label{sec:LQprops}Gauge-link configurations and light-quark propagators}

We performed the computations presented here using gauge-link configurations generated by the RBC and UKQCD collaborations \cite{Aoki:2010dy, Blum:2014tka} with 2+1 flavors of domain-wall fermions \cite{Kaplan:1992bt, Furman:1994ky, Shamir:1993zy, Brower:2012vk} and the Iwasaki gauge action \cite{Iwasaki:1984cj}. We use the five ensembles listed in Table \ref{tab:configurations}, which differ in the lattice spacing $a \approx 0.083 \, \textrm{fm} \ldots 0.114 \, \textrm{fm}$, the lattice size (spatial extent $\approx 2.65 \, \textrm{fm} \ldots 5.48 \, \textrm{fm}$) and the pion mass $m_\pi \approx 139 \, \textrm{MeV} \ldots 431 \, \textrm{MeV}$. Ensemble C00078 uses the M\"obius domain-wall action \cite{Brower:2012vk} with length of the fifth dimension $N_5=24$, while the other ensembles use the Shamir action \cite{Shamir:1993zy} with $N_5=16$. The lattice spacings listed in the Table were determined in Ref.~\cite{Blum:2014tka}.

\begin{table}[htb]
	\centering
	\begin{tabular}{ccccccccc} \hline \hline 
		Ensemble & $N_s^3 \times N_t$ & $a$ [fm] 	& $a m_{u;d}$ & $a m_{s}$ & $m_\pi$ [MeV] & $N_{\textrm{meas}}$  & $N_{\textrm{EV}}$   & $N_{\textrm{CG,sl}}$ \\ \hline
		C00078 & $48^3 \times 96$& $0.1141(3)$	& $0.00078$			      & $0.0362$           & $139(1)$		 & $\phantom{0}2560$ sl, $\phantom{0}80$  ex  & 500 & 400 \\ \hline
		C005 & $24^3 \times 64$	 & $0.1106(3)$	& $0.005\phantom{00}$ & $0.04\phantom{00}$ & $340(1)$ 	 	& $\phantom{0}9952$ sl, $311$ ex & 400 & 100 \\
		C01	 & $24^3 \times 64$	 & $0.1106(3)$  & $0.01\phantom{000}$ & $0.04\phantom{00}$ & $431(1)$		 & $\phantom{0}9056$ sl, $283$ ex  & 400 & 100  \\ \hline
		F004 & $32^3 \times 64$  & $0.0828(3)$  & $0.004\phantom{00}$ & $0.03\phantom{00}$ & $303(1)$		 & $\phantom{0}8032$ sl, $251$ ex  & 400 & 120  \\
		F006 & $32^3 \times 64$  & $0.0828(3)$	& $0.006\phantom{00}$ & $0.03\phantom{00}$ & $360(1)$		 & $14144$ sl, $442$ ex & 400 & 120  \\ \hline \hline
	\end{tabular}
	\caption{\label{tab:configurations}Gauge-link ensembles \cite{Aoki:2010dy, Blum:2014tka} and light-quark propagators used in this work. $N_s$, $N_t$: number of lattice sites in spatial and temporal directions; $a$: lattice spacing; $am_{u;d}$: bare up and down quark mass; $a m_{s}$: bare strange quark mass; $m_\pi$: pion mass. We use all-mode-averaging
\cite{Blum:2012uh,Shintani:2014vja} with 32 or 64 sloppy (sl) and 1 or 2 exact (ex) measurements per configuration; the column titled ``$N_{\textrm{meas}}$'' gives the total numbers of sloppy and exact light-quark propagators used on each ensemble. The values $N_{\textrm{EV}}$ and $N_{\textrm{CG, sl}}$ are the numbers of eigenvectors used for the deflation of the light-quark solver, and
the conjugate-gradient counts used for the sloppy propagators. }
\end{table}

Our calculation uses smeared point-to-all propagators for the up and down quarks (the smearing parameters are given in Sec.~\ref{sec:bbudInterpolators}). The computational cost of generating these propagators was reduced using the all-mode-averaging technique \cite{Blum:2012uh,Shintani:2014vja}. On each configuration, a small number of samples of ``exact'' correlation functions is combined with a large number of samples of ``sloppy'' correlation functions in such a way that the expectation value is equal to the exact expectation value, but the variance is reduced significantly due to the large number of sloppy samples \cite{Blum:2012uh,Shintani:2014vja}. The exact correlation functions are generated
from light-quark propagators computed with high precision (relative solver residual of $10^{-8}$), while the sloppy correlation functions are generated from approximate light-quark propagators. We used the conjugate gradient (CG) solver combined with low-mode deflation, where in the case of the approximate propagators the CG iteration count is fixed to a smaller value, $N_{\textrm{CG, sl}}$, than needed for the exact propagators. The lowest $N_{\textrm{EV}}$ eigenvectors of the domain-wall operator were computed using Lanczos with Chebyshev-polynomial acceleration. The values of $N_{\textrm{EV}}$ and $N_{\textrm{CG,sl}}$ are also listed in the table. On a given gauge-link configuration, the different samples were obtained by displacing the source locations on a four-dimensional grid, with a randomly chosen overall offset.

% ********************

\subsection{Bottom-quark propagators}

The heavy $b$ quarks are treated with the framework of lattice nonrelativistic QCD (NRQCD) \cite{ThackerLepage91, Lepage:1992tx}. We use the same lattice NRQCD action and parameters as in
Ref.~\cite{Brown:2014ena}. This action includes all quark-bilinear operators through order $v^4$ in the heavy-heavy power counting, and order $\Lambda^2/m_b^2$ in the heavy-light power counting.
The bare heavy-quark mass was tuned on the C005 and F004 ensembles such that the spin-averaged bottomonium kinematic mass agrees with experiment, using the lattice spacing determinations of Ref.~\cite{Meinel:2010pv}. We use the same masses also on the other coarse and fine ensembles, respectively, as shown in Table \ref{tab:NRQCDparams}. The matching coefficients $c_1$, $c_2$, $c_3$ were set to their tree-level values $(=1)$, while for $c_4$ we use a result computed at one loop in lattice perturbation theory \cite{Hammant:2011bt}. The gauge links entering the NRQCD action are divided by the mean link $u_{0L}$ in Landau gauge to achieve tadpole improvement \cite{Lepage:1992xa,Shakespeare:1998dt}.

\begin{table}[htb]
  \begin{tabular}{lcccccc}
    \hline\hline
    Ensemble & \hspace{2ex} &  $a m_b$  & \hspace{2ex} & $u_{0L}$ & \hspace{2ex} & $c_4$ \\
    \hline
    C00078                   &&  2.52  &&  0.8432  &&  1.09389 \\
    C005, C01                &&  2.52  &&  0.8439  &&  1.09389 \\
    F004, F006               &&  1.85  &&  0.8609  &&  1.07887 \\
    \hline\hline
  \end{tabular}
  \caption{\label{tab:NRQCDparams}Parameters used in the NRQCD action for the bottom quarks.}
\end{table}

Since we focus on the computation of the energy spectrum in this work, it is sufficient to use the leading-order, tree-level relation between the full-QCD bottom-quark field $b$ and the two-spinor NRQCD quark and antiquark fields $\psi$ and $\chi$ when constructing the hadron interpolating fields. In the Dirac gamma matrix basis, and omitting the phase factor that produces the tree-level energy shift, this amounts to
\begin{equation}
 b = \left( \begin{array}{c} \psi \\ \chi \end{array} \right), \quad \bar{b} = \left( \begin{array}{cc} \psi^\dag, & -\chi^\dag \end{array} \right),
\end{equation}
and the bottom-quark propagator becomes
\begin{equation}
 G_b(\xb,t,\xb',t') =  \Theta (t - t^\prime) \left(\begin{array}{cc}
G_\psi(\xb,t,\xb',t') & 0 \\
0 & 0
\end{array}\right) - \Theta (-t + t^\prime) \left(\begin{array}{cc}
0 & 0 \\
0 & G_\chi(\xb,t,\xb',t')
\end{array}\right),
\end{equation}
with the two-spinor NRQCD quark and antiquark propagators $G_\psi$ and $G_\chi$.

% ********************
% ********************
% ********************
% ********************
% ********************

\section{\label{sec:interpolators}Interpolating operators and correlation functions}

% ********************

\subsection{\label{sec:bbudInterpolators}\texorpdfstring{$\bar{b} \bar{b} u d$ four-quark system}{\bar{b} \bar{b} u d four-quark system}}

We are interested in the spectrum of the doubly-bottomed system with quantum numbers $I(J^P) = 0(1^+)$. The lowest two thresholds in this channel correspond to the meson pairs $B B^*$ and $B^*B^*$. The lowest three-meson threshold is $B B \pi$, which is approximately 44 MeV above $B^* B^*$. From Ref.~\cite{Brown:2014ena} we can see that the $\overline{\Xi}_{bb} N$ antibaryon-baryon threshold is already much higher than the $B^*B^*$ threshold: $11.1 \, \textrm{GeV}$ for $\overline{\Xi}_{bb} N$ compared to $10.6 \, \textrm{GeV}$ for $B^*B^*$.

The $J^P = 1^+$ quantum numbers appear in the  $T_1^g$ irreducible representation (irrep) of the $O_h$ point group \cite{Johnson:1982yq}. To determine the low-lying spectrum in this irrep we make use of two types of interpolating operators: local and nonlocal. The first three operators are local operators, in which all four (smeared) quark fields are multiplied at the same space-time point and the product is projected to zero momentum (in the following we omit the time coordinate):
\begin{align}
\label{eq:op_BBast_total_zero} &\op_1 = \op_{[B B^\ast](0)} = \sum_{\xb} \Big(\bar{b}(\xb) \gamma_5 d(\xb)\Big) \Big(\bar{b}(\xb) \gamma_j u(\xb)\Big) - (d \leftrightarrow u) \\
\label{eq:op_BastBast_total_zero} &\op_2 = \op_{[B^\ast B^\ast](0)} = \epsilon_{j k l} \sum_{\xb} \Big(\bar{b}(\xb) \gamma_k d(\xb)\Big) \Big(\bar{b}(\xb) \gamma_l u(\xb)\Big) - (d \leftrightarrow u) \\
\label{eq:op_Dd_total_zero} &\op_3 = \op_{[D d](0)} = \sum_{\xb} \Big(\epsilon^{a b c} \bar{b}(\xb)^b \gamma_j \C \bar{b}^{c,T}(\xb)\Big) \Big(\epsilon^{a d e} d^{d,T}(\xb) \C \gamma_5 u^e(\xb)\Big) - (d \leftrightarrow u) .
\end{align}
Operators four and five are nonlocal operators, where each color singlet is projected to zero momentum individually:
\begin{align}
\label{eq:op_BBast_each_zero} &\op_4 = \op_{B(0) B^\ast(0)} =  \bigg(\sum_{\xb} \bar{b}(\xb) \gamma_5 d(\xb)\bigg) \bigg(\sum_{\yb} \bar{b}(\yb) \gamma_j u(\yb)\bigg) - (d \leftrightarrow u) \\
\label{eq:op_BastBast_each_zero} &\op_5 = \op_{B^\ast(0) B^\ast(0)} = \epsilon_{j k l} \bigg(\sum_{\xb} \bar{b}(\xb) \gamma_k d(\xb)\bigg) \bigg(\sum_{\yb} \bar{b}(\yb) \gamma_l u(\yb)\bigg) - (d \leftrightarrow u) .
\end{align}
Above, $a,b,c,\ldots$ denote color indices, $j,k,l$ spatial vector indices, and $\C = \gamma_0 \gamma_2$ is the charge-conjugation matrix.

We expect that operators $\op_1$ to $\op_3$ generate sizable overlap to a tetraquark state. $\op_1$ is an obvious choice in studying this channel. Our reason to include $\op_2$ is a bit more subtle. Since two $B^\ast$ mesons are around $45 \, \textrm{MeV}$ heavier than a $B$ and a $B^\ast$ meson, one might expect that $\op_{[B^\ast B^\ast](0)}$ will have less overlap with the ground state than $\op_{[B B^\ast](0)}$. However, a previous investigation with static quarks (see Ref.~\cite{Bicudo:2016ooe} and Fig.~3 therein) has determined the wave functions for both $BB^*$ and $B^*B^*$ contributions and found that both spin structures are of similar importance. The inclusion of the color-triplet diquark-antidiquark interpolator $\op_3$ is motivated by the heavy-quark limit, in which the two heavy antiquarks are expected to form a compact color-triplet object, and the tetraquark becomes equivalent to a singly heavy baryon \cite{Carlson:1987hh, Manohar:1992nd, Eichten:2017ffp}. The importance of diquark operators was also discussed in Refs.~\cite{Jaffe:2004ph, Cheung:2017tnt}.

Note that in exotic channels it is typically difficult to properly resolve the ground state due to the coupling to nearby thresholds. An example of this phenomenon is the positive-parity $D_s$ spectrum, where the authors of Refs.~\cite{Mohler:2013rwa,Lang:2014yfa} had to include nonlocal interpolating operators to resolve the $D_{s0}(2317)$ mass puzzle. Previous studies of the  $I(J^P) = 0(1^+)$, bottomness-$2$ sector \cite{Francis:2016hui,Junnarkar:2018twb} did not include multi-hadron operators, which might have affected their results. To make our determination of the spectrum more robust, we include operators $\op_4$ and $\op_5$, which are nonlocal meson-meson scattering operators built from two color-singlets separately projected to zero momentum. We expect them to generate sizable overlap with the nearby first excited state, which will help us isolate the ground state in the multi-exponential fits of the correlation matrices.

To improve the overlap to the low-lying states, we employ standard smearing techniques. The quark fields in  $\op_1$ to $\op_5$ are Gaussian-smeared using
\begin{equation}
q_{\rm smeared}= \left( 1 +\frac{\sigma_\textrm{Gauss}^2}{4 N_\textrm{Gauss} }\Delta\right)^{N_\textrm{Gauss}} q, \label{eq:GaussSmearing}
\end{equation}
where $\Delta$ is the nearest-neighbor gauge-covariant spatial Laplacian. For the up and down quarks, the gauge links in $\Delta$ are spatially APE-smeared \cite{Albanese:1987ds}, while the unsmeared gauge links are used for the bottom quark. The smearing parameters are collected in Table~\ref{tab:smearingparams}.

\begin{table}[htb]
	\begin{tabular}{lccccccc} \hline \hline 
		Ensemble   & \multicolumn{4}{c}{Up and down quarks} & \hspace{2ex} & \multicolumn{2}{c}{Bottom quarks} \\
               & $N_\textrm{Gauss}$ & $\sigma_\textrm{Gauss}$ & $N_\textrm{APE}$ & $\alpha_\textrm{APE}$ && $N_\textrm{Gauss}$ & $\sigma_\textrm{Gauss}$  \\ \hline
    C00078     & $100$           & $7.171$           & $25$ & $2.5$ && $10$ & $2.0$ \\ 
    C005, C01  & $\phantom{0}30$ & $4.350$ & $25$ & $2.5$ && $10$ & $2.0$ \\
    F004, F006 & $\phantom{0}60$ & $5.728$           & $25$ & $2.5$ && $10$ & $2.0$ \\ \hline \hline
	\end{tabular}
	\caption{\label{tab:smearingparams}Parameters for the smearing of the quark fields in the interpolating operators. The Gauss smearing is defined in Eq.~(\protect\ref{eq:GaussSmearing}). A single sweep of APE smearing \cite{Albanese:1987ds} with parameter $\alpha_\textrm{APE}$ is defined as in Eq.~(8) of Ref.~\cite{Bonnet:2000dc}, and we apply $N_\textrm{APE}$ such sweeps.}
\end{table}

\begin{figure}[!htb]
  \begin{center}
  \includegraphics[width=0.90\textwidth]{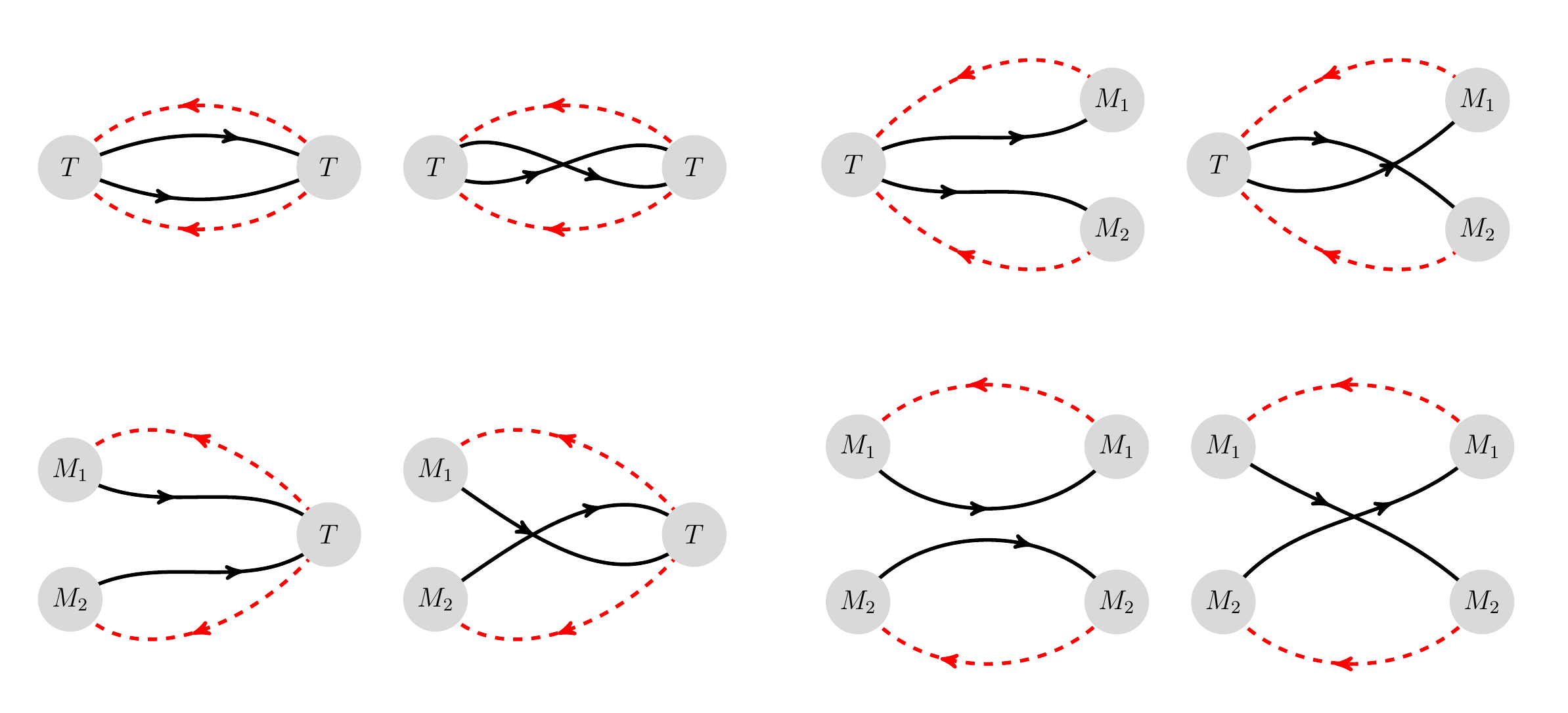}
  \caption{\label{fig:wick2pt}Nonperturbative quark-field Wick contractions for the different elements of the correlation matrix. $M_1$ and $M_2$ represent the $B$ and $B^*$ mesons that are separately projected to momentum zero, while $T$ represents the jointly projected operators. The black lines represent $b$-quark propagators and the red lines represent light-quark propagators.}
  \end{center}
\end{figure}

To determine the spectrum we compute the temporal correlation functions of the interpolating operators $\op_1$ to $\op_5$,
\begin{align}
\label{eq:cor_fct} C_{j k}(t) = \Big\langle \op_j(t) \op^\dagger_k(0) \Big\rangle ,
\end{align}
where $\langle \ldots \rangle$ denotes the path integral expectation value. The corresponding nonperturbative quark-field Wick contractions in a given gauge-field configuration are shown in Fig.~\ref{fig:wick2pt}. Because the calculation of the light-quark propagators is computationally expensive, we reuse existing smeared point-to-all propagators. With such propagators we are limited to operators $\op_1$, $\op_2$, $\op_3$ at the source, which have only a single momentum projection, allowing us to remove the summation over $\mathbf{x}$ at the source (using translational symmetry). Thus, we do not determine the elements $C_{4 4}(t)$, $C_{4 5}(t)$, $C_{5 4}(t)$ and $C_{5 5}(t)$ of the correlation matrix, where two momentum projections are needed both at the source and at the sink. For a detailed discussion of this approach, see, for example, Refs.~\cite{Abdel-Rehim:2017dok,Francis:2018qch}.

The correlation matrix has analytical properties that follow from the symmetries of lattice QCD, in particular time reversal and charge conjugation. These symmetries imply that $(C_{j k}(t))^\ast = C_{k j}(t)$ and that all $C_{j k}(t)$ are real. Moreover, one can relate $C_{j k}(+t)$ and $C_{j k}(-t)$. We exploit these analytical findings to improve our lattice QCD results, by averaging related correlation functions appropriately and by setting all imaginary parts (which are pure noise) to zero.

% ********************

\subsection{\label{sec:BandBstmeson}\texorpdfstring{$B$ and $B^\ast$ meson}{B and B* meson}}

Since we will compare the energy levels of the $\bar{b} \bar{b} u d$ four-quark system to the $B B^\ast$ threshold, we need to determine the energies of the $B$ and $B^*$ mesons within the same setup. We use the interpolating operators
\begin{align}
\label{eq:Binterpolator} & \op_B(\mathbf{p}) = \sum_{\xb} e^{i \mathbf{p} \xb} \bar{b}(\xb) \gamma_5 u(\xb), \\
\label{eq:BastInterpolator} & \op_{B^\ast}(\mathbf{p}) = \sum_{\xb} e^{i \mathbf{p} \xb} \bar{b}(\xb) \gamma_j u(\xb),
\end{align}
where we also consider nonzero momenta $\mathbf{p} = 2 \pi \mathbf{n} / L$, $\mathbf{n} \in \mathbb{Z}^3$, to allow the determination of the kinetic masses (needed for the scattering analysis in Sec.~\ref{sec:scatteringAnalysis}). The quark fields are smeared with the same parameters as in Table \ref{tab:smearingparams}. We compute the correlation functions $\langle \op_B(\mathbf{p},t) \op_B^\dagger(\mathbf{p},0) \rangle$ and $\langle \op_{B^\ast}(\mathbf{p},t) \op_{B^\ast}^\dagger(\mathbf{p},0) \rangle$. As discussed in Sec.~\ref{sec:bbudInterpolators}, we perform the summation over $\mathbf{x}$ at the sink only.

% ********************
% ********************
% ********************
% ********************
% ********************

\section{\label{sec:Bmasses}\texorpdfstring{Energies and kinetic masses of the $B$ and $B^\ast$ mesons}{Energies and kinetic masses of the B and B* mesons}}

\begin{figure}[htb]
 \includegraphics[width=0.5\linewidth]{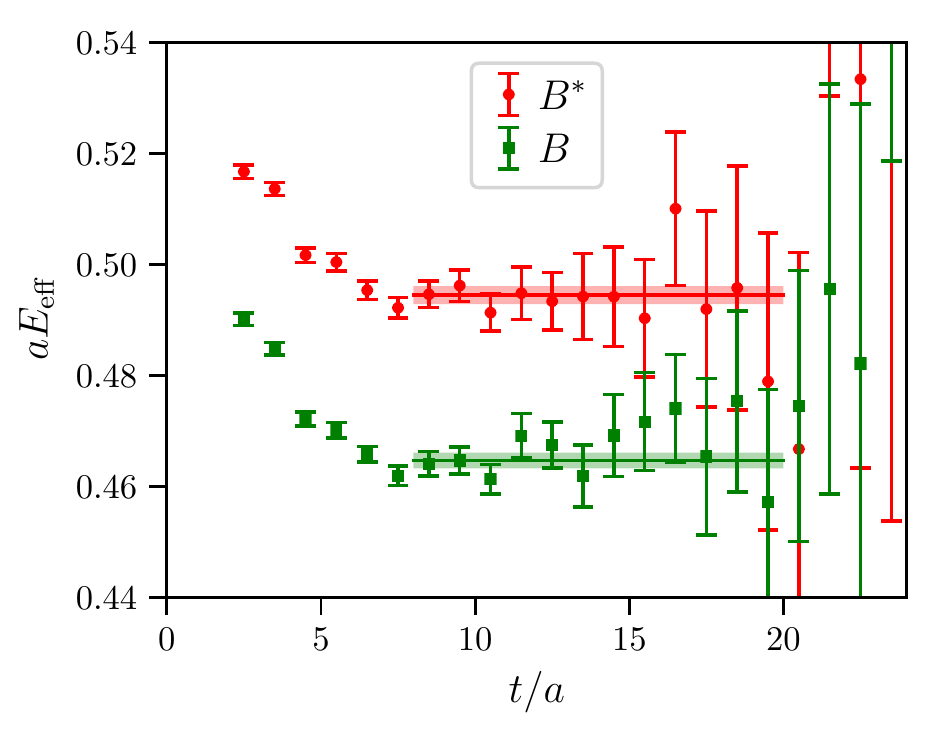}
\caption{\label{fig:BBstarEM}Effective energy plot of the $B$ and $B^*$ two-point functions at zero momentum computed on the C005 ensemble. The effective energy is defined as $aE_{\rm eff}(t+a/2)=\ln [C(t)/C(t+a)]$. The horizontal lines show the results of single-exponential fits to the two-point functions in the range $t/a=8...20$.}
\end{figure}

We determined the energies of the $B$ and $B^*$ mesons from single-exponential fits of the two-point functions; the results are listed
in Table \ref{tab:Bmasses}. An example of a corresponding effective-energy plot is shown in Fig.~\ref{fig:BBstarEM}. The energy of a state containing
$n_b$ bottom quarks is shifted by $n_b$ times the NRQCD energy shift, which at tree-level would be equal to $-m_b$. This energy shift is not known with high precision, but
cancels in energy differences with matching numbers of bottom quarks, including the quantities of interest in the following sections: $E_n - E_B - E_{B^*}$, where $E_n$ is the $n$-th energy level of the $\bar{b}\bar{b}ud$ system.

\begin{table}[htb]
  \centering
  \begin{tabular}{lcccccc} \hline \hline 
    Ensemble & $a E_B(0)$ & $a E_{B^*}(0)$ & $a m_{B,\,{\rm kin}}$ & $a m_{B^*\!,\,{\rm kin}}$ & $E_{B^*}(0)-E_B(0)$ [MeV] & $m_{\rm kin,spinav}$ [GeV] \cr 
    \hline
    C00078   & 0.4582(45)\phantom{0}   & 0.4820(46) & 3.03(14)\phantom{0}  & 3.10(13)\phantom{0}  & 41.2(5.1) & 5.33(22)\phantom{0}   \cr 
    C005     & 0.4647(14)\phantom{0}   & 0.4944(16) & 3.002(40) & 2.993(42) & 53.2(1.7) & 5.346(73)  \cr  
    C01      & 0.4742(12)\phantom{0}   & 0.5061(16) & 3.034(38) & 3.030(40) & 57.0(1.8) & 5.409(69)  \cr
    F004     & 0.3750(11)\phantom{0}   & 0.3975(12) & 2.323(21) & 2.323(25) & 51.3(1.8) & 5.536(57)  \cr
    F006     & 0.37655(87)  & 0.3985(10) & 2.320(20) & 2.311(23) & 52.3(1.5) & 5.513(54)  \cr
\hline \hline 
  \end{tabular}
  \caption{\label{tab:Bmasses}Energies of the $B$ and $B^*$ mesons at rest and kinetic masses computed using one quantum of momentum. Also given are
the hyperfine splittings and spin-averaged kinetic masses ($m_{\rm kin,spinav}=\frac14 m_{B,\,{\rm kin}} + \frac34 m_{B^*\!,\,{\rm kin}}$) in physical units. All uncertainties are statistical only.}
\end{table}

\begin{table}[htb]
  \centering
  \begin{tabular}{lcccc} \hline \hline 
    Ensemble & $c^2_B(2)$ & $c^2_B(3)$ & $c^2_{B^*}(2)$ & $c^2_{B^*}(3)$   \cr 
    \hline
    C00078   & 0.9999(13)\phantom{0} & 0.9998(26)  & 1.0007(12)\phantom{0} & 1.0014(24) \cr 
    C005     & 1.0019(14)\phantom{0} & 1.0036(28)  & 1.0014(13)\phantom{0} & 1.0026(28) \cr  
    C01      & 1.0015(12)\phantom{0} & 1.0032(25)  & 1.0000(15)\phantom{0} & 1.0002(30) \cr
    F004     & 1.00005(78)           & 1.0001(16)  & 1.00063(85)           & 1.0014(17) \cr
    F006     & 1.00032(64)           & 1.0007(13)  & 1.00030(78)           & 1.0005(16) \cr
\hline \hline 
  \end{tabular}
  \caption{\label{tab:Bspeedoflight} Results for the $B$ and $B^*$ meson ``speed of light'' squared, defined in Eq.~(\ref{eq:csqr}), for $\mathbf{p}^2 = 2 (2\pi/L)^2$ and $\mathbf{p}^2 = 3 (2\pi/L)^2$.  }
\end{table}

For the scattering analysis in Sec.~\ref{sec:scatteringAnalysis} (and to assess the tuning of the $b$-quark mass), we also need the momentum-dependence of the $B$ and $B^*$ energies. We find that within statistical uncertainties
our results are consistent with the form
\begin{eqnarray}
 E_B(\mathbf{p}) &=& E_B(0) + \sqrt{m_{B,\,{\rm kin}}^2 + \mathbf{p}^2} - m_{B,\,{\rm kin}} \label{eq:Bdisprel} \\
 E_{B^*}(\mathbf{p}) &=& E_{B^*}(0) + \sqrt{m_{B^*\!,\,{\rm kin}}^2 + \mathbf{p}^2} - m_{B^*\!,\,{\rm kin}} \label{eq:Bstardisprel}
\end{eqnarray}
up to the highest momenta we computed ($\mathbf{p}^2 = 3 (2\pi/L)^2$). The kinetic masses given in Table \ref{tab:Bmasses} were extracted using
\begin{equation}
 m_{\rm kin} = \frac{\mathbf{p}^2-[E(\mathbf{p})-E(0)]^2}{2 [E(\mathbf{p})-E(0)]}
\end{equation}
with the smallest possible non-vanishing momentum, $\mathbf{p}^2 = (2\pi/L)^2$. To test whether the $B$ and $B^*$ energies at higher momenta still satisfy the dispersion relations (\ref{eq:Bdisprel}) and (\ref{eq:Bstardisprel}) with the same parameters, we computed the square of the ``speed of light'' via 
\begin{equation}
 c^2 = \frac{[E(\mathbf{p})-E(0)+m_{\rm kin,1}]^2-m_{\rm kin,1}^2}{\mathbf{p}^2}, \label{eq:csqr}
\end{equation}
where $m_{\rm kin,1}$ are the kinetic masses computed with one unit of momentum. The results for $c^2$ are listed in Table \ref{tab:Bspeedoflight} and are found
to be consistent with 1 within their statistical uncertainties of 0.3\% or smaller.

On the C00078 and C005 ensembles, the spin-averaged kinetic masses $m_{\rm kin,spinav}=\frac14 m_{B,\,{\rm kin}} + \frac34 m_{B^*\!,\,{\rm kin}}$
agree with the experimental value of 5.28272(2) GeV \cite{Tanabashi:2018oca}. On the other ensembles, the lattice results are up to 5\% higher, which can be attributed mainly to the following:
\begin{itemize}
\item[(i)] The tuning of the bare $b$-quark mass was performed using bottomonium; the results here are affected by the heavier-than-physical light-quark masses and by discretization errors.

\item[(ii)] The tuning was performed using a different determination of the lattice spacing (that of Ref.~\cite{Meinel:2010pv}, while here we use the lattice spacing determinations of Ref.~\cite{Blum:2014tka} to convert to physical units).
\end{itemize}
However, the effect of a possible $\lesssim 5 \%$ mistuning of the $b$-quark mass is expected to be even smaller in the energy differences $E_n - E_B - E_{B^*}$ due to partial suppression by heavy-quark symmetry. The hyperfine splittings $E_{B^*} - E_B$ are of order $\Lambda/m_b$, and are therefore affected by the same relative error as the $b$-quark mass,  which corresponds to an absolute error of $\lesssim 2.5 \, \textrm{MeV}$. Our results for the hyperfine splittings
are also shown in Table \ref{tab:Bmasses}. The value from the physical-pion-mass ensemble C00078 is consistent with the experimental value of $45.3(2)\: \textrm{MeV}$ \cite{Tanabashi:2018oca}. We find a trend of increasing hyperfine splitting as the light-quark mass is increased.

% ********************
% ********************
% ********************
% ********************
% ********************

\section{\label{sec:bbudEnergies}\texorpdfstring{The lowest energy levels of the $\bar{b} \bar{b} u d$ system}{The lowest energy levels of the \bar{b} \bar{b} u d system}}

% ********************

\subsection{\label{sec:multiExpFitting}Multi-exponential matrix fitting}

\begin{figure}[htb]
  \centering
   \includegraphics[width=0.6\linewidth]{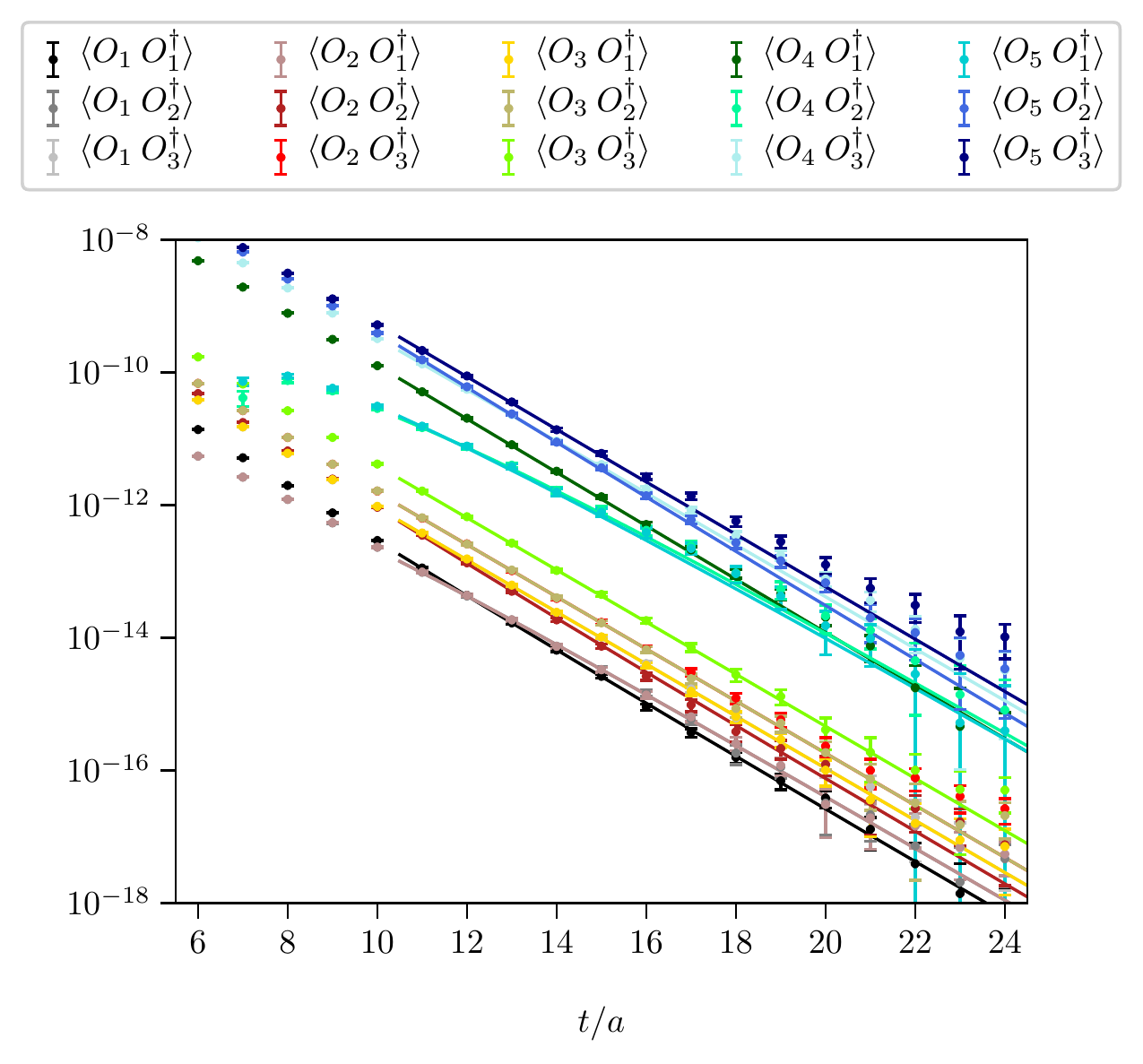}
  \caption{\label{fig:MFAexample}Example of a multi-exponential matrix fit for the C005 ensemble. This fit has $N=3$, $t_{\rm min}/a=11$, $t_{\rm max}/a=24$, and gives $\chi^2/{\rm d.o.f.}\approx 1.08$.}
\end{figure}

The spectral decomposition of the correlation matrix (\ref{eq:cor_fct}) reads
\begin{align}
C_{j k}(t) = \sum_{n=0}^{\infty}\langle \Omega |\op_j | n \rangle \langle  n | \op^\dagger_k | \Omega \rangle \textrm{e}^{-E_n t} , \label{eq:cor_spec_decomp}
\end{align}
where $| \Omega \rangle$ denotes the vacuum state, $| n \rangle$ are the energy eigenstates of the $\bar{b}\bar{b}ud$ system in the $T_1^g$ irrep and for isospin $I = 0$, and $E_n$ are the corresponding energy eigenvalues. Equation (\ref{eq:cor_spec_decomp}) assumes an infinite time extent of the lattice, which is a good approximation in our case. As discussed in Sec.~\ref{sec:bbudInterpolators}, all $ C_{j k}(t) $ are real and, thus, so are the overlap factors
\begin{equation}
Z_j^n = \langle \Omega |\op_j | n \rangle.
\end{equation}
To extract the energies $E_n$ from our numerical results for $C_{j k}(t)$, we perform fully correlated, least-$\chi^2$, multi-exponential fits using a truncated version of (\ref{eq:cor_spec_decomp}),
\begin{align}
\label{eq:multiexp} C_{j k}(t) \approx \sum_{n=0}^{N-1} Z_j^n Z_k^n \textrm{e}^{-E_n t},
\end{align}
where we must choose the time range $t_\textrm{min} \leq t \leq t_\textrm{max}$ such that contributions from higher excited states are negligible. To enforce the ordering
of the $E_n$ returned from the fit, for $n>0$ we actually use the logarithms of the energy differences, $l_n = \ln(aE_n - aE_{n-1})$, as our fit parameters. Furthermore, we
rewrote the overlap factors for $n>0$ as $Z_n = B_n Z_0$ and used $B_n$ as the fit parameters. Note that $C_{j k}(t)$ does not need to be a square matrix. In particular, the multi-exponential matrix fit method allows us to also include correlation matrix elements $C_{j k}(t)$ with $j = 4,5$ and $k = 1,2,3$, i.e.\ the scattering operators $\op_4$ and $\op_5$. An example of such a $(5\times 3)$ matrix fit is shown in Fig.~\ref{fig:MFAexample}. On each ensemble, we performed fits to the full matrix as well as to various sub-matrices, for different fit ranges and different numbers of states. The results are listed in Tables \ref{tab:MFitresultsC00078}-\ref{tab:MFitresultsF006} in Appendix~\ref{app:energytablesandfigures}.

The matrix fits have a large number of degrees of freedom, and the covariance matrix, whose inverse enters in $\chi^2$, can become poorly determined or singular if the number of statistical samples used to estimate this matrix is not much larger than the number of degrees of freedom. This is particularly problematic when using all-mode-averaging (or, more generally, covariant-approximation-averaging, CAA), where the samples are given by \cite{Blum:2012uh,Shintani:2014vja}
\begin{equation}
 (\text{CAA sample})_e = (\text{exact sample})_e - (\text{sloppy sample})_{e,0} + \frac{1}{N_{\text{sloppy}}} \sum_{s=0}^{N_{\text{sloppy}}-1} (\text{sloppy sample})_{e,s}.
\end{equation}
Here, $e$ labels the exact samples (gauge-link configuration and source location), and the different sloppy samples, computed from quark propagators with reduced solver precision, in our case originate from applying many space-time displacements $s$ to the initial source location $e$, with $s=0$ corresponding to no displacement. As shown in Table~\ref{tab:configurations}, the number of exact samples, and hence CAA samples, is as low as 80 in the case of the C00078 ensemble. To obtain a meaningful $\chi^2$ even for large numbers of degrees of freedom, we use a ``modified CAA'' (MCAA) procedure, given by

\begin{equation}
 (\text{MCAA sample})_{e,s} = (\text{exact sample})_e - (\text{sloppy sample})_{e,0} + (\text{sloppy sample})_{e,s}.
\end{equation}
This procedure provides $N_{\text{sloppy}}$ $(=32$ in our case) times as many samples as the standard CAA procedure, without changing the overall average, and allows robust matrix fits even in the $(5\times 3)$ case. The drawback is that there are autocorrelations between the different choices of $s$, introduced by the constant (but very small) term $(\text{exact sample})_e - (\text{sloppy sample})_{e,0}$ and by any possible autocorrelations between the different sloppy samples on the same configuration. As a result of these autocorrelations, the uncertainties of parameters obtained from fits based on the MCAA procedure are initially slightly underestimated. We correct for these autocorrelations by rescaling all uncertainties in the fitted $E_n$ with a factor estimated for each ensemble using the simple $B$ meson two-point functions. The factor is given by the ratio of uncertainties of the $B$ meson energies obtained from single-exponential fits using the CAA and MCAA procedures. The factors range from 1.08 to 1.27. The uncertainties of all results shown in this paper are already corrected with these factors.

The condition numbers of the data covariance matrices range from approximately $10^6$ to $10^{15}$, depending on how many interpolating operators are included and on the fit range. We always computed the full inverse using LU decomposition. The minimization of $\chi^2$ was performed using the Levenberg-Marquardt algorithm, with initial guesses for the fit parameters obtained as described in Sec.~V B of Ref.~\cite{Alexandrou:2017mpi}. The fit results were stable under variations of the initial guesses within reasonable ranges.

As a cross-check of the multi-exponential matrix fitting method, we also determined the spectrum using the variational approach, which involves solving the generalized eigenvalue problem (GEVP) \cite{Luscher:1990ck,Blossier:2009kd,Orginos:2015tha}. We found that the GEVP method, where applicable, gives results consistent with the direct multi-exponential matrix fits.
The comparison of the two methods is presented in Appendix~\ref{app:comparisonGEVPmultiexp}.

% ********************

\FloatBarrier
\subsection{\label{sec:OperatorDependence}Dependence of the fit results on the choices of interpolating fields }

\begin{figure}[htb]
  \centering
  \includegraphics[width=0.7\textwidth]{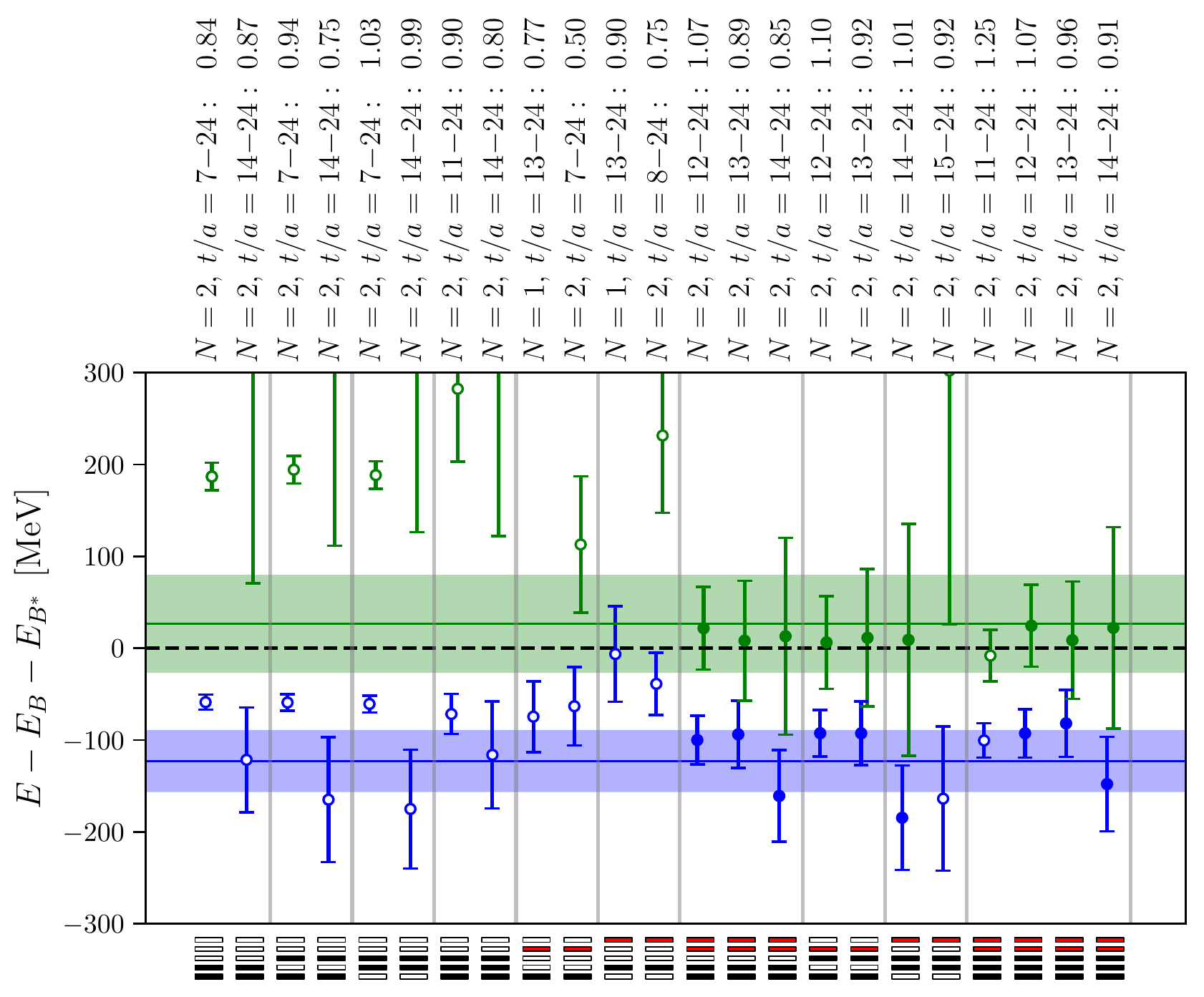}
  \caption{\label{fig:SpectrumC005}Results for the lowest two $\bar{b}\bar{b}ud$ energy levels relative to the $B B^\ast$ threshold, $\Delta E_n = E_n - E_B - E_{B^*}$, as determined on ensemble C005 from several different fits. The five bars below each column indicate the interpolators used, as explained in the main text. Above each column, we give the number of exponentials, the fit range, and the value of $\chi^2/\textrm{d.o.f.}$. The shaded horizontal bands correspond to our final estimates of $\Delta E_0$ and $\Delta E_1$, obtained from a bootstrap average of the subset of fits that are shown with filled symbols.}
\end{figure}

Even though the actual energy levels for a chosen set of quantum numbers are independent of the interpolating operators used in the two-point functions, in practice the numerical results
depend on these choices, due to limited statistical precision. In Fig.~\ref{fig:SpectrumC005} we present the two lowest energy levels relative to the $B B^\ast$ threshold as determined on ensemble C005 using multi-exponential matrix fitting. The interpolators used are indicated by the five bars below each column. The three black bars at the bottom correspond to the local interpolators $\op_1$, $\op_2$, $\op_3$, while the two red bars at the top correspond to the nonlocal interpolators $\op_4$ and $\op_5$. A filled bar indicates that an interpolator is included, an empty bar that it is not included. Two things are evident from Fig.~\ref{fig:SpectrumC005}: first, stable results for the two lowest energy levels are only obtained once the nonlocal two-meson interpolators are included in the analysis; second, once the nonlocal interpolators are included, the estimates of both the ground-state energy and the first excited energy drop significantly. Both observations clearly indicate the importance of the nonlocal interpolators for our study of the $\bar{b} \bar{b} u d$ system. Note that the first excited energy is close to the $B B^\ast$ threshold, while the ground state energy is significantly below threshold. This is a first indication that the ground state corresponds to a stable tetraquark, while the first excitation corresponds to a meson-meson scattering state. Similar plots for the other four ensembles are collected in Appendix~\ref{app:energytablesandfigures}.

% ********************

\FloatBarrier
\subsection{\label{sec:OverlapFactors}Overlap factors}

For a given $j$, the overlap factors $Z_j^n$ indicate the relative importance of the energy eigenstates $| n \rangle$ when expanding the trial state $\op_j^\dagger | \Omega \rangle$ in terms of energy eigenstates,
\begin{align}
\op_j^\dagger | \Omega \rangle = \sum_{n=0}^\infty | n \rangle \langle n | \mathcal{O}_j^\dagger | \Omega \rangle = \sum_{n=0}^\infty Z_j^n | n \rangle .
\end{align}
Therefore, the overlap factors $Z_j^n$ provide certain information about the composition and quark arrangement of the energy eigenstates $| n \rangle$. In particular, if the overlap factor $Z_j^m$ for one state $|m\rangle$ is significantly larger than all other $Z_j^n$, $n \neq m$, this might be a sign that $| m \rangle$ is quite similar to $\op_j^\dagger | \Omega \rangle$.

It is convenient to consider rescaled squared overlap factors,
\begin{align}
\label{eq:Zrat} |\tilde{Z}_j^n|^2 = \frac{|Z_j^n|^2}{\textrm{max}_m(|Z_j^m|^2)} ,
\end{align}
which are normalized such that $\textrm{max}_m(|\tilde{Z}_j^m|^2) = 1$ for each trial state $\op_j^\dagger | \Omega \rangle$. Here, the indices $n$ and $m$ can take on values from 0 to $N-1$, where $N$ is the number of states included in the fit. The results for $|\tilde{Z}_j^n|^2$ obtained from our fits are qualitatively similar for all ensembles, and do not strongly depend on the temporal fit range $t_\textrm{min} \leq t \leq t_\textrm{max}$.

\begin{figure}[b]
  \centering
  \includegraphics[width=0.8\textwidth]{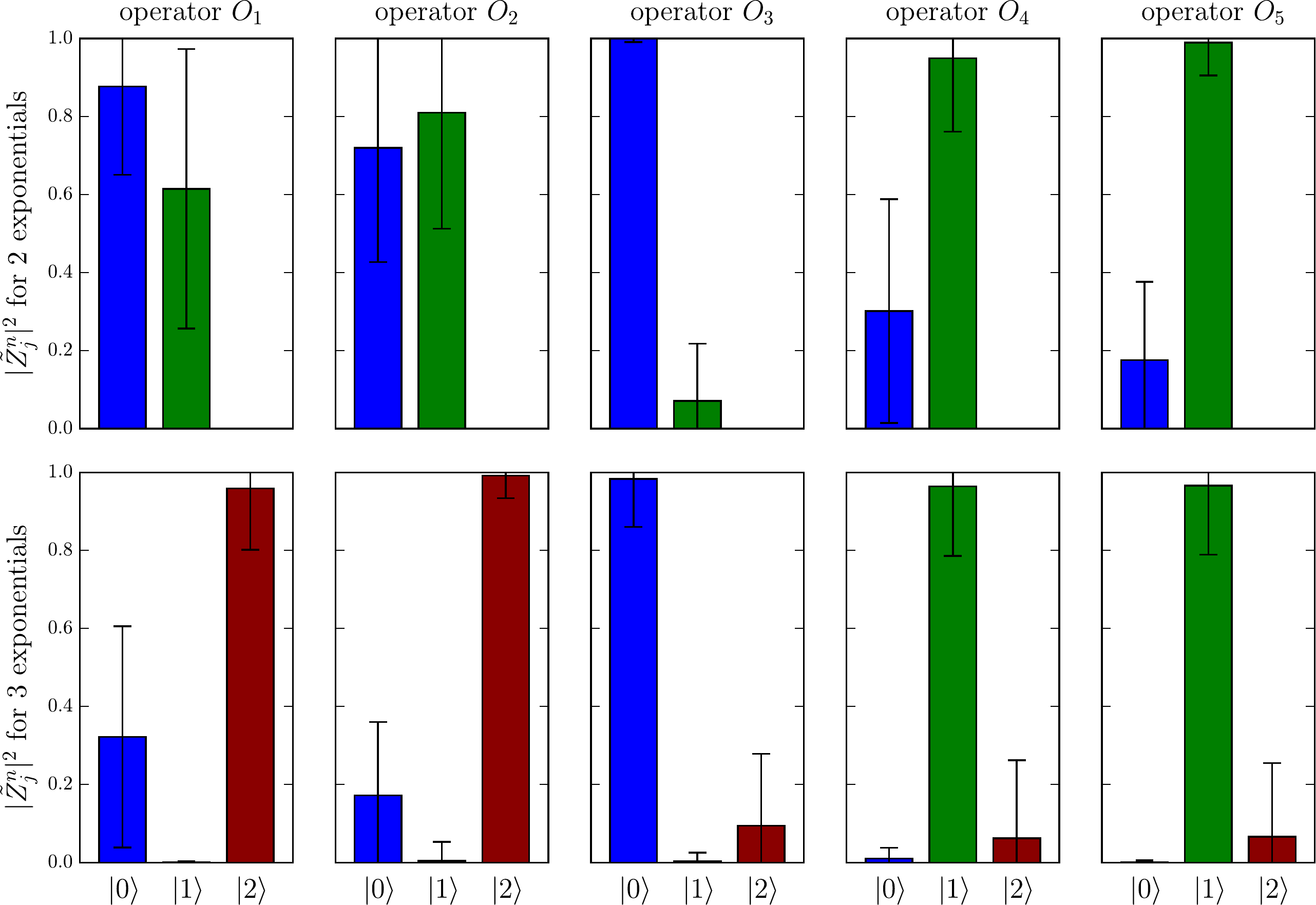}
  \caption{\label{fig:overlaps}The normalized overlap factors $|\tilde{Z}_j^n|^2$ as determined on ensemble C005, indicating the relative contributions of the energy eigenstates $| n \rangle$ to the trial state $\mathcal{O}_j^\dagger | \Omega \rangle$. The upper row corresponds to a two-exponential fit with $14 \leq t/a \leq 24$, while the lower row corresponds to a three-exponential fit with $12 \leq t/a \leq 24$. The mean values and variances of the $|\tilde{Z}_j^n|^2$ were evaluated using bootstrap resampling of the fits, which explains why the mean of the largest factor can be smaller than 1.}
\end{figure}

In Fig.~\ref{fig:overlaps} we show the normalized overlap factors obtained on ensemble C005 using the full $5 \times 3$ correlation matrix for $N = 2$ and $N = 3$, with fit ranges of $14 \leq t/a \leq 24$ and $12 \leq t/a \leq 24$, respectively. One can see that for the trial state created by the diquark-antidiquark operator $\op_3$, the ground-state overlap is significantly larger than the overlap to the first excited state. Vice versa, for the two trial states created by the nonlocal meson-meson operators $\op_4$ and $\op_5$, the overlaps to the first excited state are larger than the ground-state overlaps. This supports our above interpretation concerning the composition of the lowest two energy eigenstates: the ground state $| 0 \rangle$ seems to be a four-quark bound state and the first excitation $| 1 \rangle$ a meson-meson scattering state.
Finally, it is interesting to note that in the three-exponential fit, the local meson-meson operators $O_1$ and $O_2$ appear to produce a large overlap with the \emph{second} excited state, while the diquark-antidiquark operator $\op_3$ and the nonlocal operators $\op_4$ and $\op_5$ do not (also note that the large values of $|Z_1^2|^2/|Z_1^0|^2=4\pm5$, $|Z_2^2|^2/|Z_2^0|^2=8\pm9$ substantially change the normalization of the plots for $O_1$ and $O_2$ when going from $N=2$ to $N=3$). What appears in the fit as the ``second excited state'' (with a rather high energy and large uncertainty, as shown in Table \ref{tab:MFitresultsC005}) is likely an admixture of the dense spectrum of all the scattering states above threshold, which are all created with similar weights by the local operators $O_1$ and $O_2$, while the nonlocal operators $O_4$ and $O_5$ mostly create the lowest-lying scattering state.

% ********************

\FloatBarrier
\subsection{\label{sec:finitevolumeFinalenergies}Final results for the lowest two energy levels on each ensemble}

To obtain final results for the energy differences $\Delta E_0$ and $\Delta E_1$ on each ensemble, we select a subset of fits that we deem reliable, based on the discussion
in the previous two subsections and based on $\chi^2/{\rm d.o.f.}$. All of these fits, which are indicated with filled symbols in Fig.~\ref{fig:SpectrumC005} and Figs.~\ref{fig:SpectrumC00078}-\ref{fig:SpectrumF006}, include at least one of the nonlocal interpolating operators $\op_4$ and $\op_5$. We then repeat all selected fits for 500
bootstrap samples, where each bootstrap sample consists of randomly drawn gauge-link configurations and source locations (using the same random numbers for the different types of fits to preserve correlations). This produces 500 samples for $\Delta E_0$ and $\Delta E_1$ for each type of fit (and on each ensemble). We then average these values over the different types of fits with equal weights, bootstrap sample by bootstrap sample. Finally, we compute the mean and standard deviation (rescaled again by the MCAA autocorrelation correction factors; see the discussion in Sec.~\ref{sec:multiExpFitting}) of these new 500 samples. The results for $\Delta E_0$ and $\Delta E_1$ obtained in this way are listed in Table \ref{tab:bbudFVresults} and are indicated with the horizontal lines and
shaded uncertainty bands in Fig.~\ref{fig:SpectrumC005} and Figs.~\ref{fig:SpectrumC00078}-\ref{fig:SpectrumF006}.

\begin{table}[htb]
  \centering
  \begin{tabular}{lcccc} \hline \hline 
    Ensemble & $a \Delta E_0$ & $a \Delta E_1$ & $\Delta E_0$ [MeV] & $\Delta E_1$ [MeV] \cr 
    \hline
    C00078   & $-0.075(16)$ & $\phantom{+}0.001(21)$   & $-129(27)$ & $\phantom{+0}2(36)$     \cr 
    C005     & $-0.069(19)$ & $\phantom{+}0.015(30)$   & $-123(34)$ & $\phantom{+}27(53)$    \cr  
    C01      & $-0.061(15)$ & $-0.014(22)$  & $-109(26)$ & $-26(39)$   \cr
    F004     & $-0.051(11)$ & $-0.011(15)$  & $-122(25)$ & $-25(37)$   \cr
    F006     & $-0.037(12)$ & $\phantom{+}0.025(32)$   & $\phantom{0}$$-88(29)$  & $\phantom{+}61(76)$    \cr
\hline \hline 
  \end{tabular}
  \caption{\label{tab:bbudFVresults}Final results for the lowest two energy levels of the $\bar{b}\bar{b}ud$ system on each ensemble, in lattice units and in MeV. All results are given relative to the $B B^*$ threshold, i.e.\ $\Delta E_n = E_n - E_B - E_{B^*}$.}
\end{table}

% ********************
% ********************
% ********************
% ********************
% ********************

\FloatBarrier

\section{\label{sec:scatteringAnalysis}Scattering analysis}
  
The energies $E_n$ determined in a spectroscopy computation can be related to the infinite-volume scattering amplitude using L\"uscher's method \cite{Luscher:1990ux} and its generalizations \cite{Rummukainen:1995vs,Kim:2005gf,Christ:2005gi,Hansen:2012tf,Leskovec:2012gb,Gockeler:2012yj,Briceno:2014oea,Briceno:2017tce}; see  Ref.\ \cite{Briceno:2017max} for a recent review.
Here, we apply L\"uscher's method to the lowest two energy levels of the $\bar{b}\bar{b}ud$ system in the $T_1^g$ irrep, assuming that these energy levels can be described in terms of the elastic $S$-wave $B$-$B^*$ scattering amplitude (and its analytic continuation below threshold). We neglect higher partial waves and the coupling to other channels, such as $B^*B^*$. Thus, we obtain the $S$-wave $B$-$B^*$ scattering amplitude for the two scattering momenta $k_0$ and $k_1$ that correspond to $E_0$ and $E_1$. Having only these two points limits the choice of parametrization of the scattering amplitude to a function with two parameters, for which we use the effective-range expansion (ERE). The ERE parametrization then allows us to determine the infinite-volume bound-state energy.

% ********************

\subsection{Relation between finite-volume energy levels and infinite-volume phase shifts}

We define the scattering momentum $k_n$ corresponding to the $n$-th energy level of the $\bar{b} \bar{b} u d$ system $E_n$   through the equation
\begin{equation}
 E_n = E_B + \sqrt{m_{B,\,{\rm kin}}^2 + k_n^2} - m_{B,\,{\rm kin}} + E_{B^*} + \sqrt{m_{B^*\!,\,{\rm kin}}^2 + k_n^2} - m_{B^*\!,\,{\rm kin}},
\end{equation}
where $E_B=E_B(0)$ and $E_{B^*}=E_{B^*}(0)$ are the energies of the single-$B$ meson and single-$B^*$ meson states at zero momentum. Solving for $k_n^2$ gives
\begin{equation}
 k_n^2 = \frac{\Delta E_n (\Delta E_n+2 m_{B,\,{\rm kin}}) (\Delta E_n+2 m_{B^*\!,\,{\rm kin}}) (\Delta E_n+2 m_{B,\,{\rm kin}}+2 m_{B^*\!,\,{\rm kin}})}{4 (\Delta E_n+m_{B,\,{\rm kin}}+m_{B^*\!,\,{\rm kin}})^2}, \label{eq:knsqr}
\end{equation}
where
\begin{equation}
 \Delta E_n = E_n-E_B-E_{B^*}.
\end{equation}

The mapping between the finite-volume scattering momentum in the rest frame and the infinite-volume scattering amplitude expressed as the $S$-wave scattering phase shift $\delta_0$ is
\begin{align}
\label{eq:Luscher} \cot{\delta_0(k_n)} = \frac{2 Z_{00}(1;(k_n L/2 \pi)^2)}{\pi^{1/2} k_n L} ,
\end{align}
where $Z_{00}$ is the generalized zeta function \cite{Luscher:1990ux}. The scattering amplitude is given by
\begin{align}
T_0(k) = \frac{1}{\cot\delta_0(k) - i}. \label{eq:scatteringamp}
\end{align}
%

% ********************

\subsection{Effective-range expansion and determination of the bound-state pole}

On each ensemble, we use the lattice QCD results for the energy differences $\Delta E_0$ and $\Delta E_1$ combined with the corresponding results for the $B$ and $B^*$ meson energies and kinetic
masses to calculate $k_n^2$ for $n=0$ and $n=1$ using Eq.~(\ref{eq:knsqr}), and we determine the corresponding $k_n \cot\delta_0(k_n)$ using Eq.~(\ref{eq:Luscher}). We parametrize the scattering amplitude using the effective-range expansion (ERE),
\begin{align}
\label{eq:effective_range} k \cot \delta_0(k) = \frac{1}{a_0} + \frac{1}{2} r_0 k^2 + \mathcal{O}(k^4) ,
\end{align}
and determine the two parameters $a_0$ (the $S$-wave scattering length) and $r_0$ (the $S$-wave effective range). Bound states correspond to poles in the scattering amplitude (\ref{eq:scatteringamp}) below threshold, where $-ik>0$. Such a pole occurs for
$\cot\delta_0(k_{\rm BS}) = i$, where $k_{\rm BS}$ is the (imaginary) bound-state momentum. Combining this condition with the ERE then gives
\begin{align}
\label{eq:BS} -|k_{\rm BS}| = \frac{1}{a_0} - \frac{1}{2} r_0 |k_{\rm BS}|^2 ,
\end{align}
where terms of $\mathcal{O}(|k_{\rm BS}|^4)$ are neglected. We solve Eq.~(\ref{eq:BS}) for $|k_{\rm BS}|$ and obtain the binding energy via
\begin{align}
E_{\rm binding} = E_{\rm BS}-E_B - E_{B^*} &= \sqrt{m_{B,\,{\rm kin}}^2 + k_{\rm BS}^2} - m_{B,\,{\rm kin}} + \sqrt{m_{B^*\!,\,{\rm kin}}^2 + k_{\rm BS}^2} - m_{B^*\!,\,{\rm kin}} \cr
 &= \sqrt{m_{B,\,{\rm kin}}^2 - |k_{\rm BS}|^2} - m_{B,\,{\rm kin}} + \sqrt{m_{B^*\!,\,{\rm kin}}^2 - |k_{\rm BS}|^2} - m_{B^*\!,\,{\rm kin}}.
\end{align}

This approach was previously utilized in Refs.~\cite{Beane:2010em,Mohler:2013rwa,Lang:2014yfa,Moir:2016srx}. Determining masses of bound states in this way is equivalent to taking the infinite-volume limit up to exponentially small finite-volume corrections proportional to $e^{-|k_{\rm BS}| L}$ (for more details, see, for example, Refs.~\cite{Sasaki:2006jn,Beane:2011iw}).

\begin{figure}
  \centering
  \includegraphics[height=0.2\textheight]{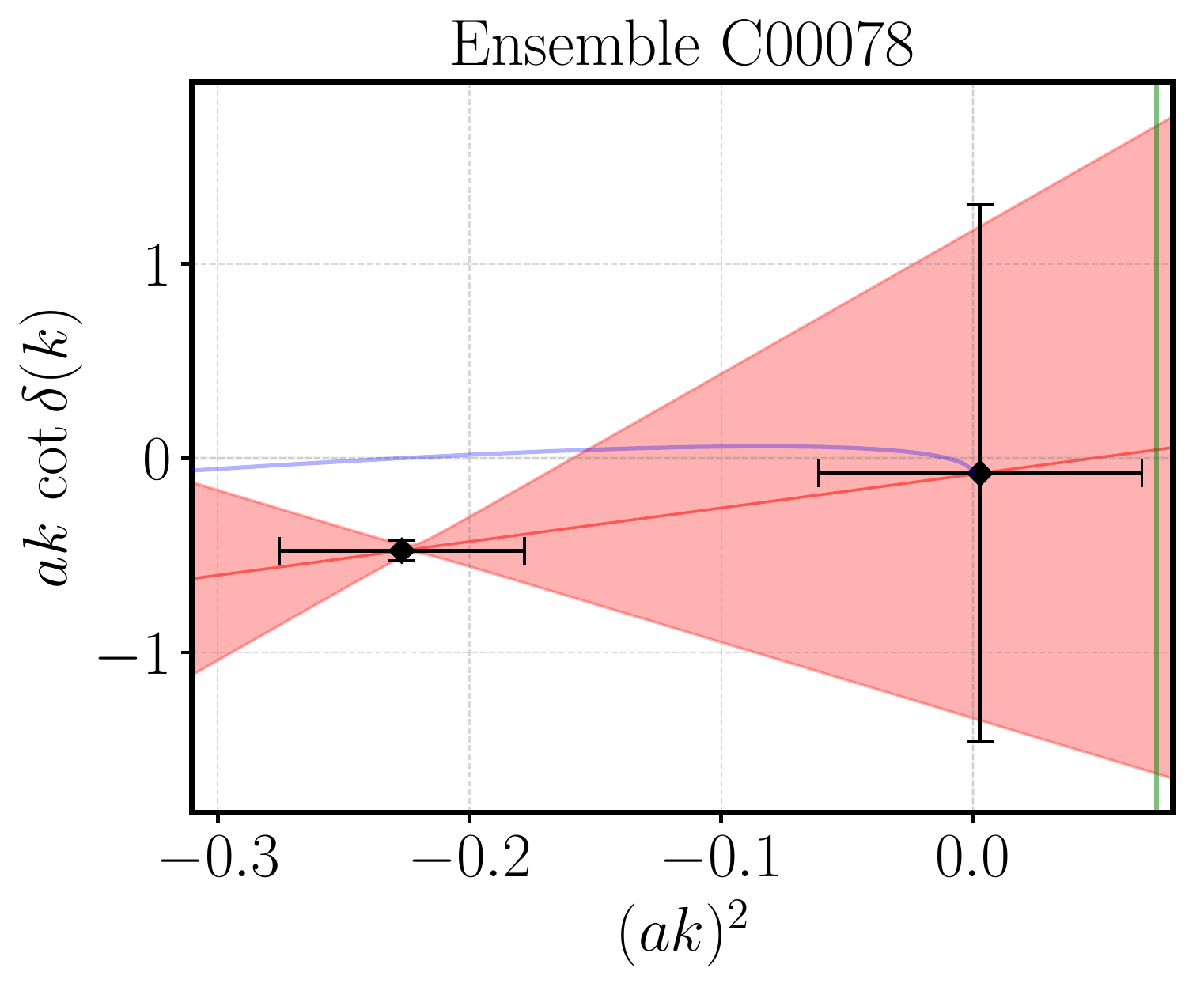}
  \includegraphics[height=0.2\textheight]{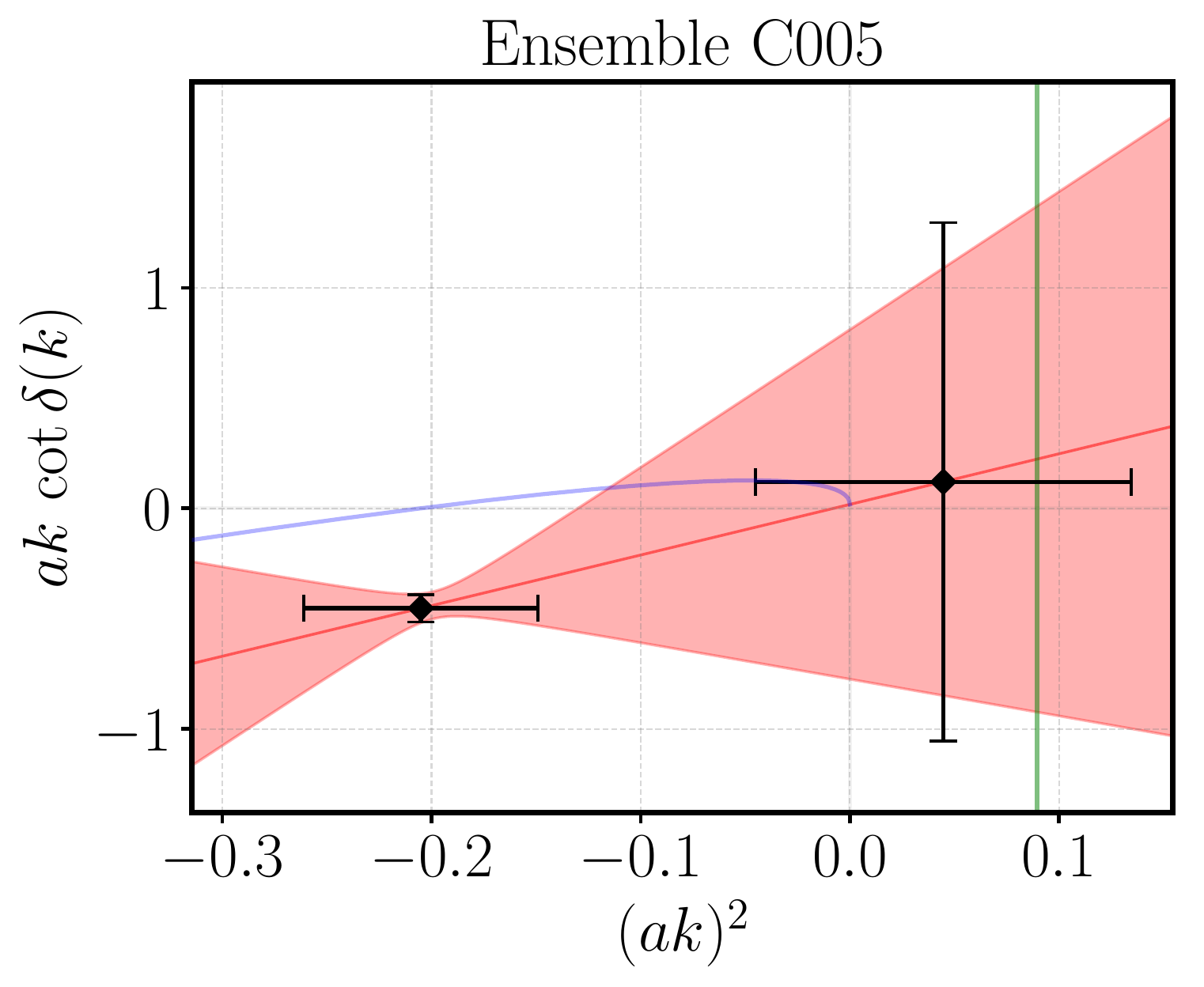}
  \includegraphics[height=0.2\textheight]{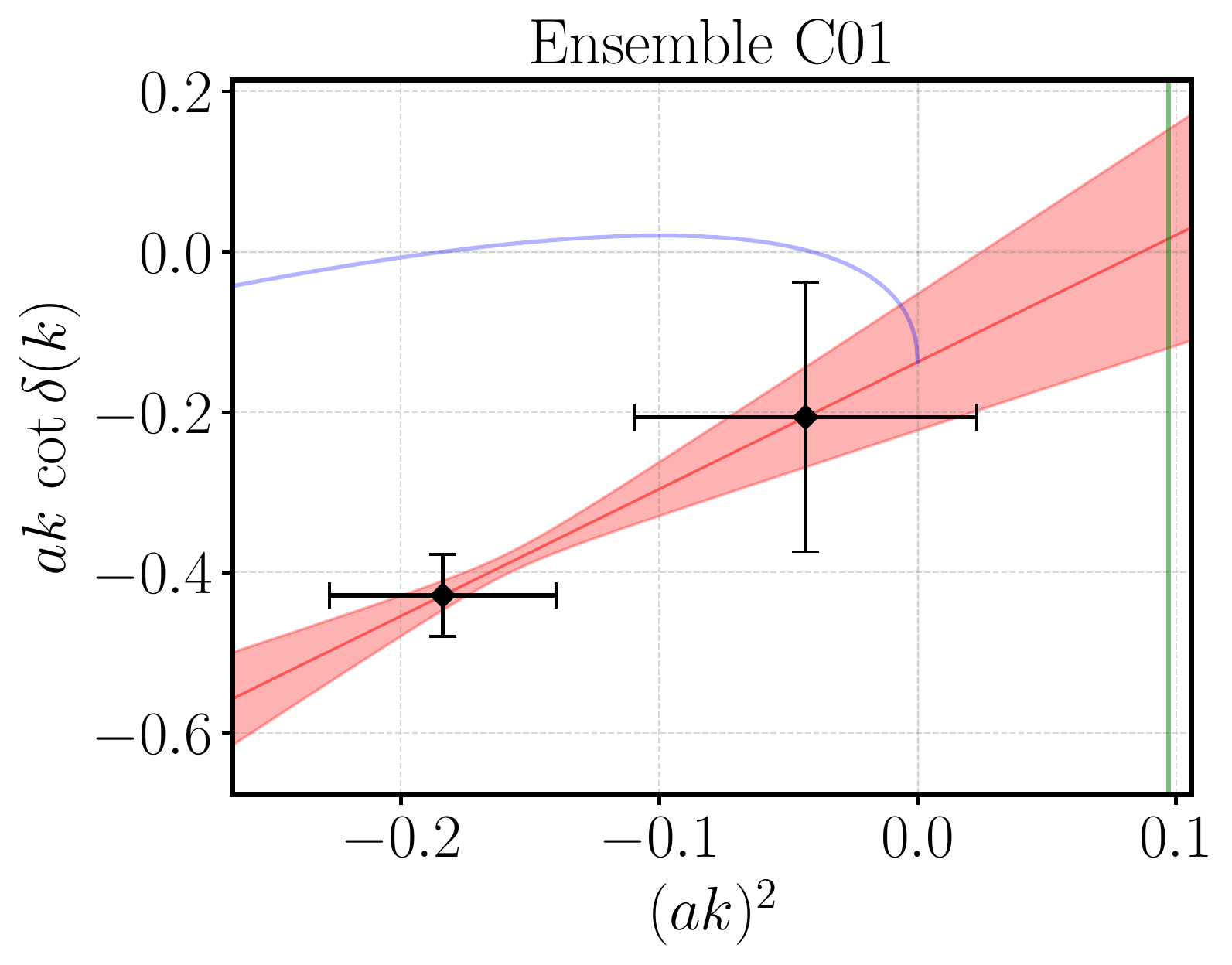} \\
  \includegraphics[height=0.2\textheight]{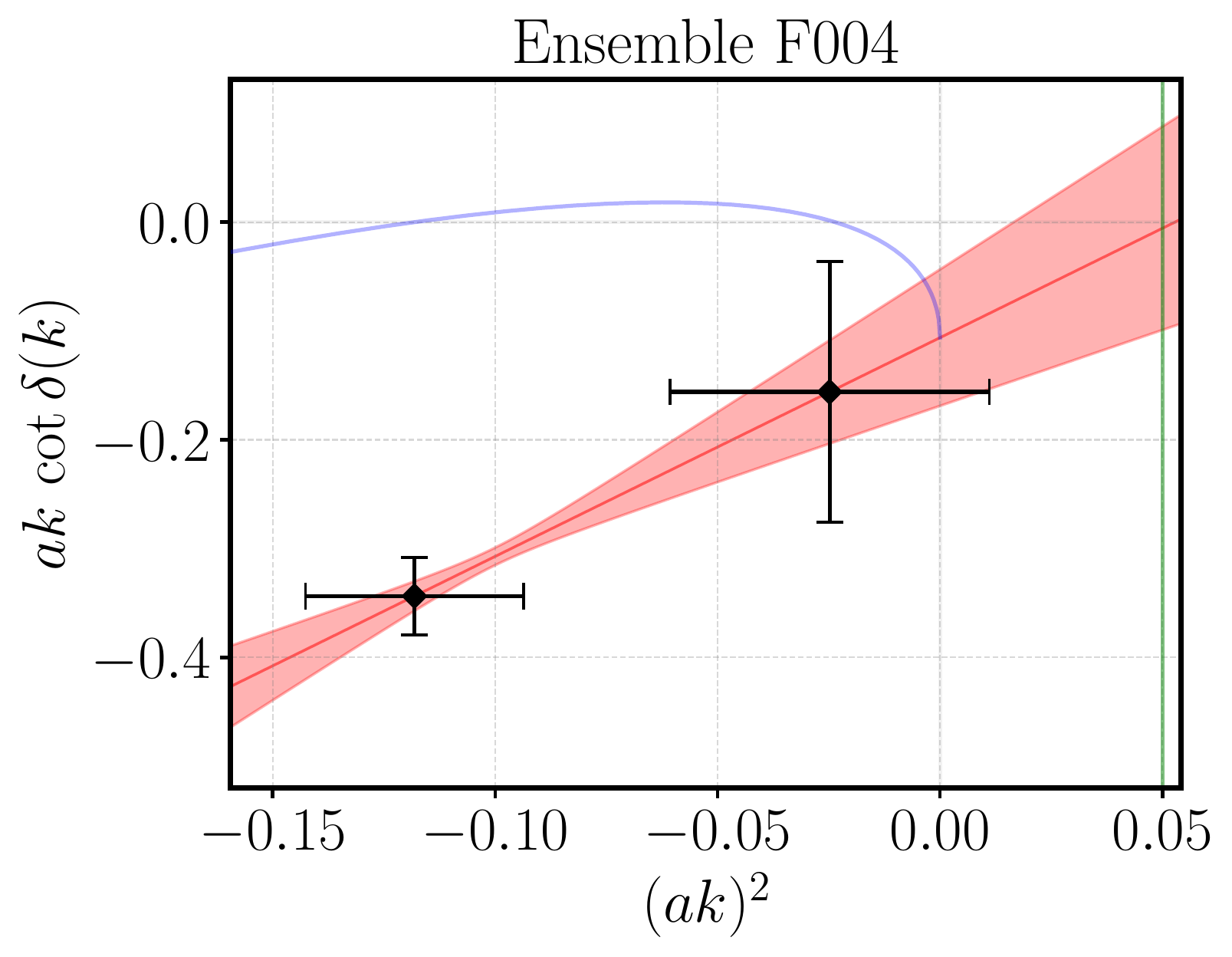}
  \includegraphics[height=0.2\textheight]{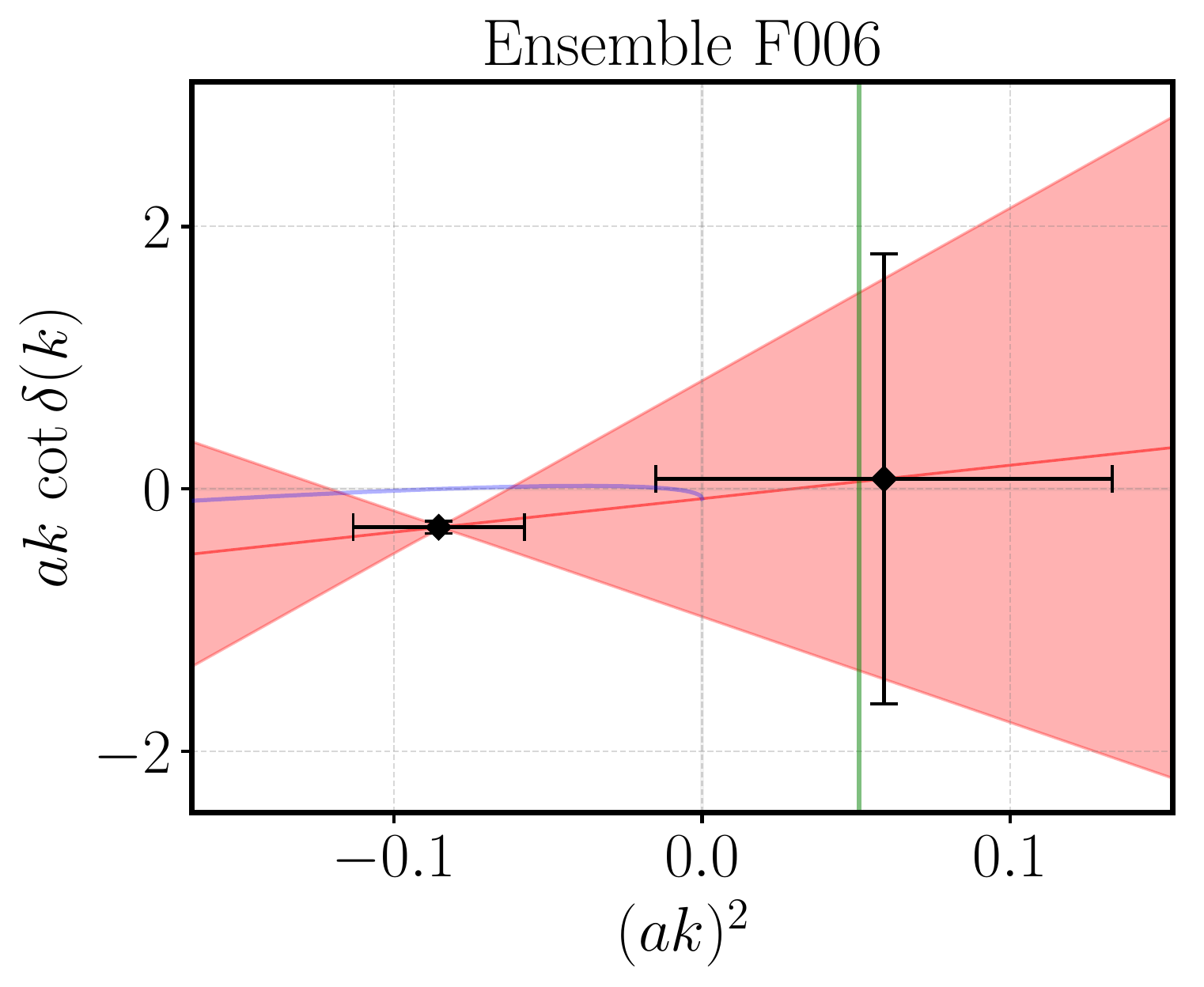}
  \caption{\label{fig:EREplot}Plots of the effective-range-expansions.  Here, $a$ denotes the lattice spacing. The straight red line going through the data points corresponds to the ERE parametrization of $ak \cot \delta(k)$. Below the $B B^*$ threshold (which is located at $k=0$), we also show the curves $ak \cot \delta(k) + |ak|$ [again using the ERE parametrization for $ak \cot \delta(k)$], whose lowest zero gives the binding momentum. The vertical green lines show the location of the inelastic $B^* B^*$ threshold.  }
\end{figure}

% ********************

\subsection{\label{sec:ScatteringResults}Numerical results}

In Fig.~\ref{fig:EREplot} we show $k \cot \delta_0(k) = 1/a_0 + r_0 k^2 / 2$ as a function of the scattering momentum $k$ together with the corresponding lattice data points, for the five different ensembles. One can see that $k \cot \delta_0(k)$ is consistent with zero within uncertainties at $k = 0$, which implies an inverse scattering length also consistent with zero.
The results for the inverse scattering length $1/a_0$, the effective range $r_0$, the binding momentum $|k_{\rm BS}|$, and the binding energy $E_\textrm{binding}$ are given in Table \ref{tab:scatt_results}. The binding energies are essentially identical to the finite-volume energy differences $\Delta E_0$ collected in Table~\ref{tab:bbudFVresults}. This supports our interpretation of the ground state as a stable tetraquark. We also applied the consistency check proposed in Ref.~\cite{Iritani:2017rlk}, but our statistical
uncertainties are too large to reach a definitive conclusion.

As indicated in Fig.~\ref{fig:EREplot}, the first excited state is close to the inelastic $B^*B^*$ threshold for some of the data sets, which means that the single-channel analysis performed here is not well justified. A coupled-channel analysis is not feasible with the data we have, but is unlikely to significantly affect the results for the bound state.

\begin{table}[h]
\centering
\begin{tabular}{l|cccc}
\hline\hline
Ensemble & $1/a_0$ [fm$^{-1}$] & $r_0$ [fm] & $|k_{\rm BS}|$ [MeV]  & $E_\textrm{binding}$ [MeV] \cr
\hline
% 1/a_0
C00078   & $-0.7(11.0)$    & $ 0.4(1.3)\phantom{.} $  & $ 824(89)\phantom{0} $  & $ -129(28) $   \cr
C005     & $\phantom{+}0.2(7.2)\phantom{0}$    & $ 0.51(88) $  & $ 808(110) $ & $ -123(34) $   \cr
C01      & $-1.24(77)\phantom{.}$ & $ 0.35(12) $  & $ 765(91)\phantom{0} $  & $ -109(26) $   \cr
F004     & $-1.28(76)\phantom{.}$ & $ 0.33(10) $  & $ 819(85)\phantom{0} $  & $ -122(25) $   \cr
F006     & $-0.9(10.9)$    & $ 0.4(1.8)\phantom{.} $  & $ 697(113)$  & $ -88(29)\phantom{0} $   \cr
\hline\hline
\end{tabular}
\caption{\label{tab:scatt_results}The inverse scattering length $1/a_0$, the effective range $r_0$, the binding momentum $|k_{\rm BS}|$, and the binding energy $E_\textrm{binding}$ for all ensembles.}
\end{table} 

% ********************
% ********************
% ********************
% ********************
% ********************

\FloatBarrier

\section{\label{sec:errors}Fit of the pion-mass dependence and estimates of systematic uncertainties}

\begin{figure}[t]
  \centering
  \includegraphics[width=0.5\textwidth]{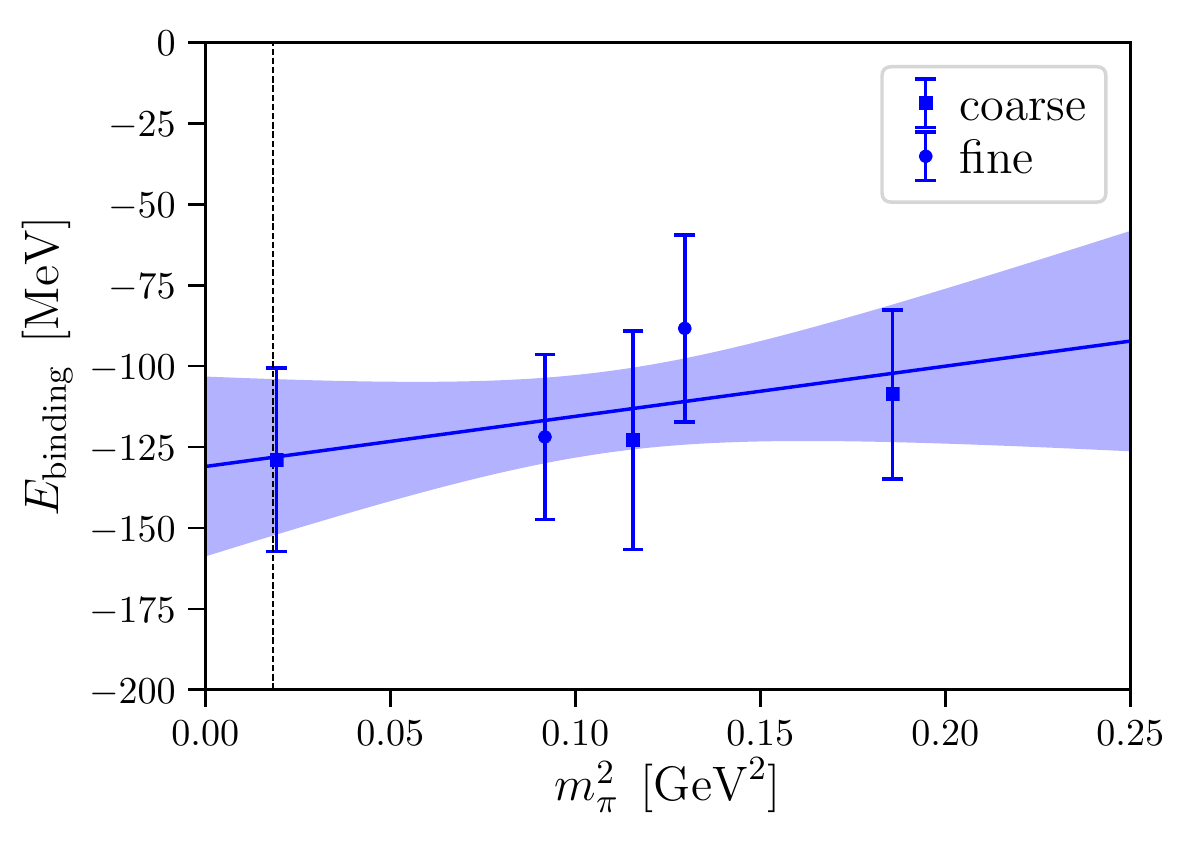}
  \caption{\label{fig:chiralExtrapolation}Fit of the pion-mass dependence of $E_\textrm{binding}$. The vertical dashed line indicates the physical pion mass.}
\end{figure}

Given the statistical uncertainties in our results for $E_\textrm{binding}$ (shown in Table~\ref{tab:scatt_results}), we cannot resolve any significant dependence on the lattice spacing or pion mass.
We expect lattice discretization errors to be at the level of a few MeV, as discussed further below. Since this is well below our statistical uncertainties, we choose to perform
a fit of the pion-mass dependence of our results from all ensembles without including $a$-dependence. We consider a quadratic pion-mass dependence, corresponding to linear dependence on the light-quark mass,
\begin{align}
\label{eq:chiralExtrap} E_\textrm{binding}(m_\pi) = E_\textrm{binding}(m_{\pi,\text{phys}}) + c (m_\pi^2 - m_{\pi,\text{phys}}^2),
\end{align}
where we use $m_{\pi,\text{phys}} = 135 \, \textrm{MeV}$ for the physical pion mass in the isospin-symmetric limit. The fit gives
\begin{eqnarray}
E_\textrm{binding}(m_{\pi,\text{phys}}) = (-128 \pm 24) \, \textrm{MeV}  , \quad\quad c = (1.5 \pm 2.3) \times 10^{-4} \, \textrm{MeV}^{-2},
\end{eqnarray}
and has $\chi^2/{\rm d.o.f.}=0.27$. A plot of the fit function together with the data is shown in Fig.~\ref{fig:chiralExtrapolation}. Given that, (i), the fit has excellent quality,
(ii), the resulting coefficient $c$ is consistent with zero, and (iii), the result for $E_\textrm{binding}(m_{\pi,\text{phys}})$ is nearly identical with the result from the C00078 ensemble with $m_\pi=139(1)$ MeV,
we conclude that any remaining systematic uncertainties associated with the extrapolation to $m_{\pi,\text{phys}}$ are negligible.

The lattice discretization errors associated with our light-quark and gluon actions and are expected to be at the 1\% level for the fine lattices,
and at the 2\% level for the coarse lattices \cite{Blum:2014tka}; multiplying by the QCD scale of $\Lambda \sim 300 \, \textrm{MeV}$ yields $3 \, \textrm{MeV}$ to $6 \, \textrm{MeV}$. The
NRQCD action introduces additional systematic uncertainties. For our choice of lattice discretization and matching coefficients, the most significant systematic errors in the energy of a heavy-light hadron are expected to be the following:
\begin{itemize}
 \item Four-quark operators, which arise at order $\alpha_s^2$ in the matching to full QCD, are not included in the action. The analysis of Ref.~\cite{Blok:1996iz} suggests that their effects could be as large as $3 \, \textrm{MeV}$.
 \item The matching coefficient $c_4$ of the operator $-\frac{g}{2 m_b}\:\bss{\sigma}\cdot\mathbf{B}$ was computed to one loop. Missing higher-order corrections to this coefficient
       introduce systematic errors of order
         \begin{equation}
           \alpha_s^2 \Lambda^2 / m_b \approx 2 \, {\rm MeV}.
         \end{equation}
 \item The matching coefficients of the operators of order $(\Lambda/m_b)^2$ were computed at tree-level only. The missing radiative corrections to these coefficients introduce
        systematic errors of order
           \begin{equation}
             \alpha_s \Lambda^3 / m_b^2 \approx 0.4 \, {\rm MeV}.
           \end{equation}
\end{itemize}
These estimates are appropriate for $E_B$ and $E_{B^*}$, which contribute to our calculation of the binding energy via $E_{\rm binding} = E_{\rm BS}-E_B - E_{B^*}$. For the tetraquark energy $E_{\rm BS}$,
the power counting is more complicated due to the presence of two bottom quarks. Conservative estimates of the systematic errors can be obtained by replacing the scale $\Lambda$ in the heavy-light power counting
by the binding momentum $|k_{\rm BS}|\sim 800$ MeV, which suggests systematic errors of order 10 MeV. It is likely that there is a partial cancellation of the systematic errors in $E_{\rm binding} = E_{\rm BS}-E_B - E_{B^*}$.
Therefore, we estimate the overall discretization and heavy-quark systematic errors to be not larger than $10 \, \textrm{MeV}$ (in future work, the estimates of the heavy-quark errors could be made more precise by numerically investigating the dependence of $E_\textrm{binding}$ on the lattice NRQCD matching coefficients). Our final results for the tetraquark binding energy and mass are therefore
\begin{eqnarray}
E_\textrm{binding}(m_{\pi,\text{phys}}) = (-128 \pm 24 \pm 10) \, \textrm{MeV} , \quad\quad m_\textrm{tetraquark}(m_{\pi,\text{phys}}) = (10476 \pm 24 \pm 10) \, \textrm{MeV} , \label{eq:finalresult}
\end{eqnarray}
where $m_\textrm{tetraquark}$ is obtained by adding the experimental values of the $B$ and $B^*$ masses \cite{Tanabashi:2018oca} to $E_\textrm{binding}$.

% ********************
% ********************
% ********************
% ********************
% ********************

\section{\label{sec:conclusions}Conclusions}

In this work we computed the low-lying spectrum in the bottomness-$2$ and $I(J^P) = 0(1^+)$ sector. Using both local and nonlocal interpolating operators, we
determined the two lowest energy levels for five different ensembles of lattice gauge-link configurations, including one with approximately physical light-quark masses.
We carried out a L\"uscher analysis for the first time in this sector and used the effective-range expansion to find the infinite-volume binding energies (but we found these to be nearly identical to the finite-volume binding energies). Our calculation confirms the existence of a $\bar{b} \bar{b} u d$ bound state that is stable under the strong and electromagnetic interactions.

\begin{figure}
  \begin{center}
  \includegraphics[width=0.95\textwidth]{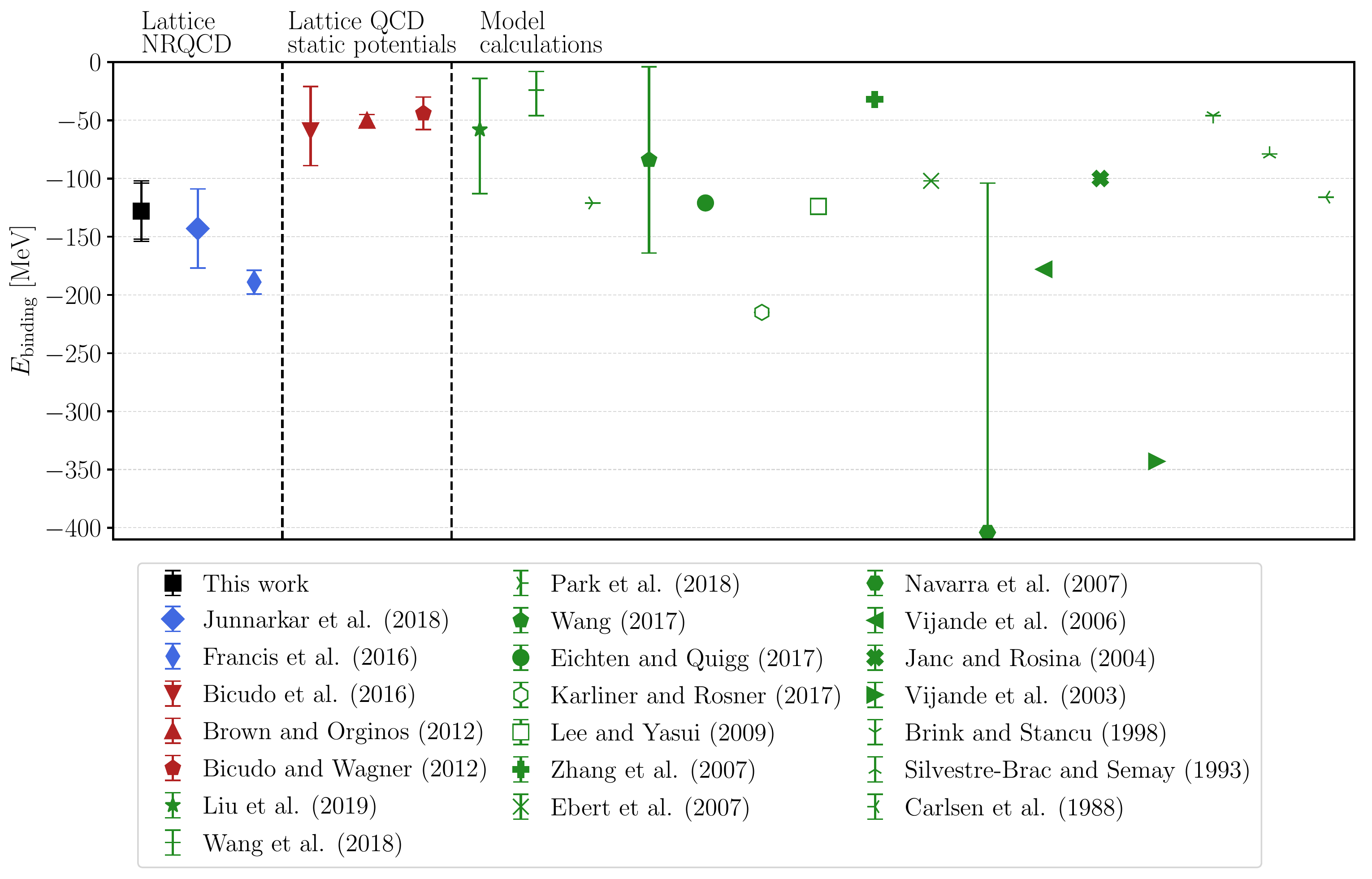}
  \caption{\label{fig:Ecompare}Comparison of results for the binding energy of the $\bar{b} \bar{b} u d$ tetraquark with $I(J^P) = 0(1^+)$ (black: this work, using lattice NRQCD; blue: previous work using lattice NRQCD  \cite{Francis:2016hui,Junnarkar:2018twb}; red: lattice QCD computations of static $\bar{b} \bar{b}$ potentials and solving the Schr\"odinger equation \cite{Bicudo:2012qt,Brown:2012tm,Bicudo:2016ooe}; green: quark models, effective field theories, and QCD sum rules \cite{Carlson:1987hh, SilvestreBrac:1993ss, Brink:1998as, Vijande:2003ki, Janc:2004qn, Vijande:2006jf, Navarra:2007yw, Ebert:2007rn, Zhang:2007mu, Lee:2009rt, Karliner:2017qjm, Eichten:2017ffp, Wang:2017uld, Park:2018wjk, Wang:2018atz, Liu:2019stu}.}
  \end{center}
\end{figure}

In Fig.~\ref{fig:Ecompare} we compare our result (\ref{eq:finalresult}) for the binding energy with several previous determinations. These include direct lattice QCD calculations that also treated the heavy $\bar{b}$ quarks with NRQCD \cite{Francis:2016hui,Junnarkar:2018twb}, calculations in the Born-Oppenheimer approximation using static $\bar{b} \bar{b}$ potentials (in the presence of $u$ and $d$ valence quarks) computed on the lattice \cite{Bicudo:2012qt,Brown:2012tm,Bicudo:2016ooe}, as well as the studies of Refs.~\cite{Carlson:1987hh, SilvestreBrac:1993ss, Brink:1998as, Vijande:2003ki, Janc:2004qn, Vijande:2006jf, Navarra:2007yw, Ebert:2007rn, Zhang:2007mu, Lee:2009rt, Karliner:2017qjm, Eichten:2017ffp, Wang:2017uld, Park:2018wjk, Wang:2018atz, Liu:2019stu}, which are based on quark models,  effective field theories, or QCD sum rules.

The calculations using static $\bar{b} \bar{b}$ potentials from lattice QCD \cite{Bicudo:2012qt,Brown:2012tm,Bicudo:2016ooe} consistently give a binding energy that is about a factor of 2 smaller than our result (\ref{eq:finalresult}), but this disagreement might be due the approximations used there, in particular the neglect of $1/m_b$ corrections to the potentials.

The two previous direct lattice QCD calculations \cite{Francis:2016hui,Junnarkar:2018twb} employed only local four-quark interpolating operators. According to our observations, the lack of nonlocal operators can affect the reliability of the extracted ground-state energy, as the local operators are not well suited to isolate the lowest $B B^*$ threshold state. While the result of Ref.~\cite{Junnarkar:2018twb} agrees with ours, Ref.~\cite{Francis:2016hui} gives a significantly larger binding energy. Apart from the lack of nonlocal interpolating operators, another possible source of this discrepancy might be the use of ratios of correlation functions as input to the generalized eigenvalue problem in Ref.~\cite{Francis:2016hui}. It is interesting to observe that the effective energies shown in Refs.~\cite{Francis:2016hui, Junnarkar:2018twb} approach the ground state from below, corresponding to a decrease in the magnitude of the extracted binding energy as the time separation is increased. Both of these studies used wall sources for the quark fields, while we use Gaussian-smeared sources. These two types of sources can behave quite differently with regard to excited-state contamination \cite{Yamazaki:2017jfh}. Furthermore, the excited-state spectrum of $B$-$B^*$ and $B^*$-$B^*$ scattering states above threshold is very dense, because the changes in the kinetic energy when increasing the back-to-back momenta are suppressed by the heavy-meson masses. For example, on a lattice with $L=6\:{\rm fm}$, the energy difference between the threshold and next scattering state is only around $8 \, \textrm{MeV}$. In the context of two-nucleon systems, it has been argued that the dense spectrum can lead to ``fake plateaus'' at short time separations in the effective energies from ratios of correlation functions \cite{Iritani:2016jie}; see Ref.~\cite{Davoudi:2017ddj} for a critical discussion of this issue.

Even though our work has improved upon previous studies of the $\bar{b}\bar{b}ud$ system by including nonlocal meson-meson scattering operators at the sink, our fits still require rather large time separations, leading to large statistical uncertainties. As demonstrated for the case of the $H$ dibaryon in Ref.~\cite{Francis:2018qch}, the results can be vastly improved by including nonlocal operators at both source and sink, and by including additional back-to-back momenta to map out a larger region of the spectrum. This requires more advanced techniques \cite{Peardon:2009gh, Abdel-Rehim:2017dok} for constructing the correlation functions.

% ********************
% ********************
% ********************
% ********************
% ********************

\FloatBarrier

\section*{Acknowledgements}

We thank Antje Peters for collaboration in the early stages of this project, and we thank the RBC and UKQCD collaborations for providing the gauge-field ensembles.
L.L.\ is supported by the U.S. Department of Energy, Office of Science, Office of Nuclear Physics under contract DE-AC05-06OR23177.
S.M.\ is supported by the U.S. Department of Energy, Office of Science, Office of High Energy Physics under Award Number D{E-S}{C0}009913, and by the RHIC Physics Fellow Program of the RIKEN BNL Research Center.
M.W.\ acknowledges support by the Heisenberg Programme of the DFG (German Research Foundation), grant WA 3000/3-1.
This work was supported in part by the Helmholtz International Center for FAIR within the framework of the LOEWE program launched by the State of Hesse. 
Calculations on the LOEWE-CSC and on the on the FUCHS-CSC high-performance computer of the Frankfurt University were conducted for this research. We would like to thank HPC-Hessen, funded by the State Ministry of Higher Education, Research and the Arts, for programming advice.
This research used resources of the National Energy Research Scientific Computing Center (NERSC), a U.S.\ Department of Energy Office of Science User Facility operated under Contract No.\ DE-AC02-05CH11231. This work also used resources at the Texas Advanced Computing Center that are part of the Extreme Science and Engineering Discovery Environment (XSEDE), which is supported by National Science Foundation grant number ACI-1548562.

% ********************
% ********************
% ********************
% ********************
% ********************

\appendix

\section{\label{app:energytablesandfigures}The two lowest energy levels for all ensembles}

\begin{figure}[!htb]
  \centering
  \includegraphics[width=0.65\textwidth]{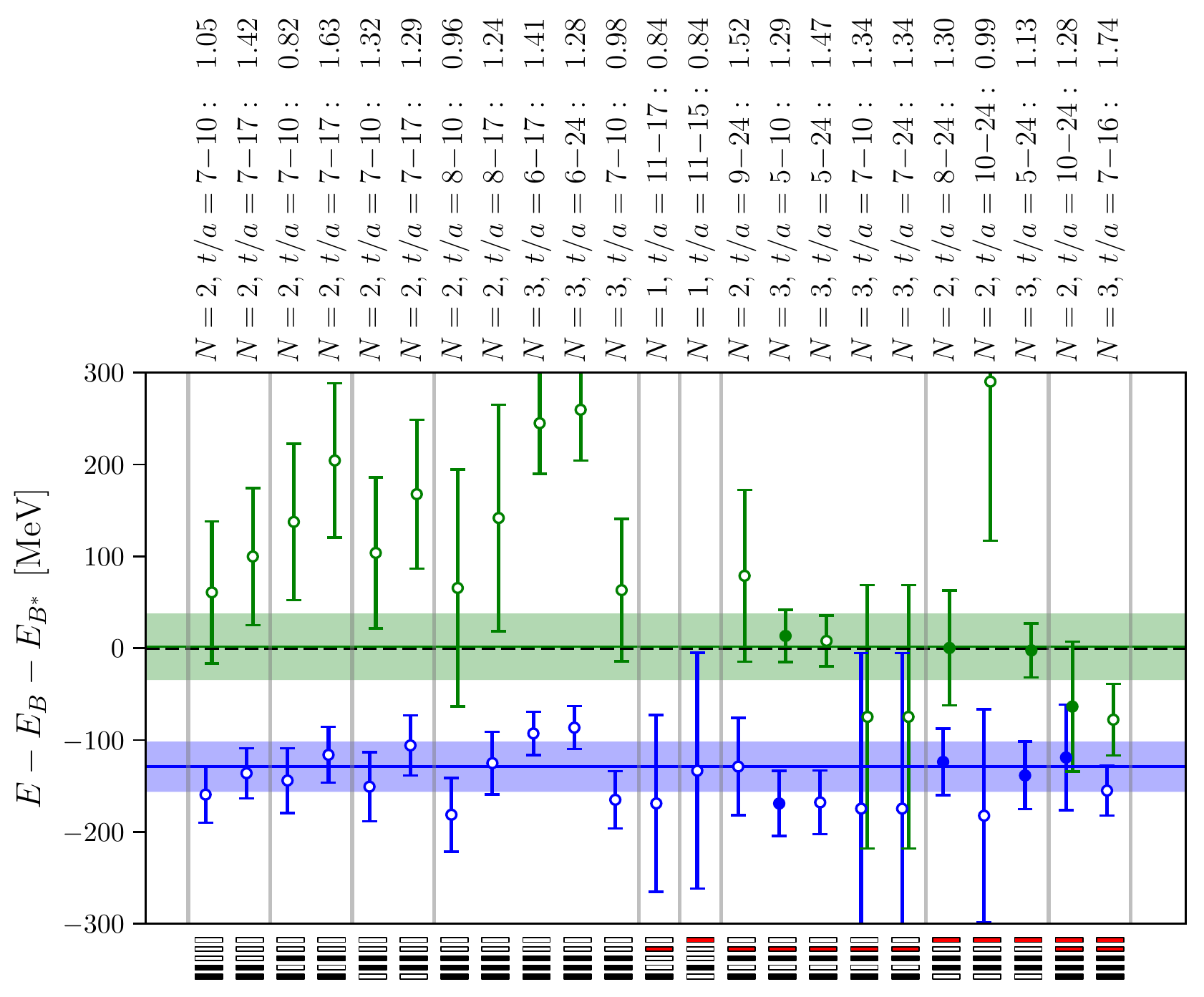}

  \vspace{-2ex}

  \caption{\label{fig:SpectrumC00078} Like Fig.~\protect\ref{fig:SpectrumC005}, but for the C00078 ensemble.}

  \vspace{-2ex}

\end{figure}

\begin{figure}[!htb]
  \centering
  \includegraphics[width=0.65\textwidth]{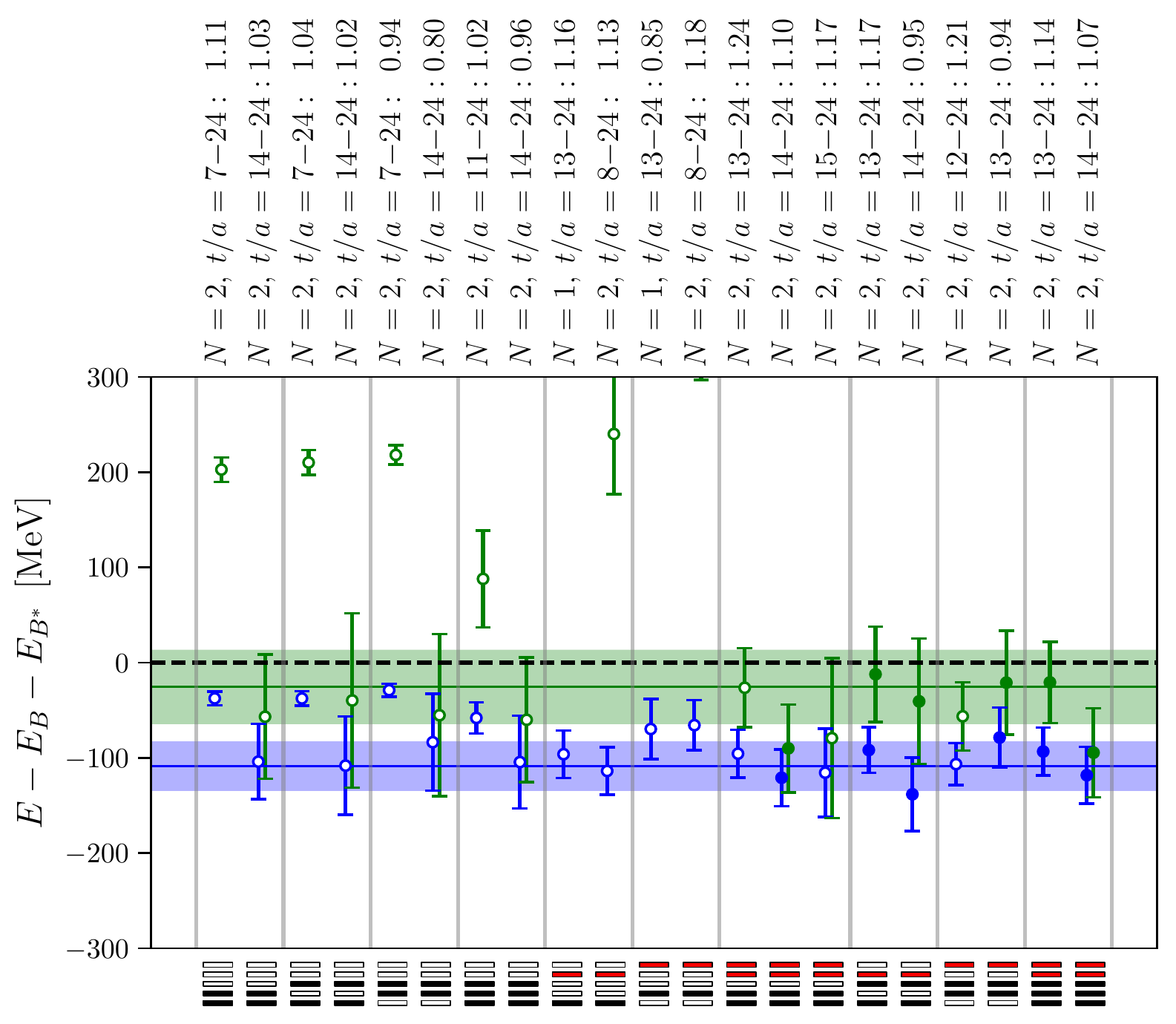}

  \vspace{-2ex}

  \caption{\label{fig:SpectrumC01} Like Fig.~\protect\ref{fig:SpectrumC005}, but for the C01 ensemble.}

  \vspace{-4ex}

\end{figure}

\begin{figure}[!htb]
  \centering
  \includegraphics[width=0.65\textwidth]{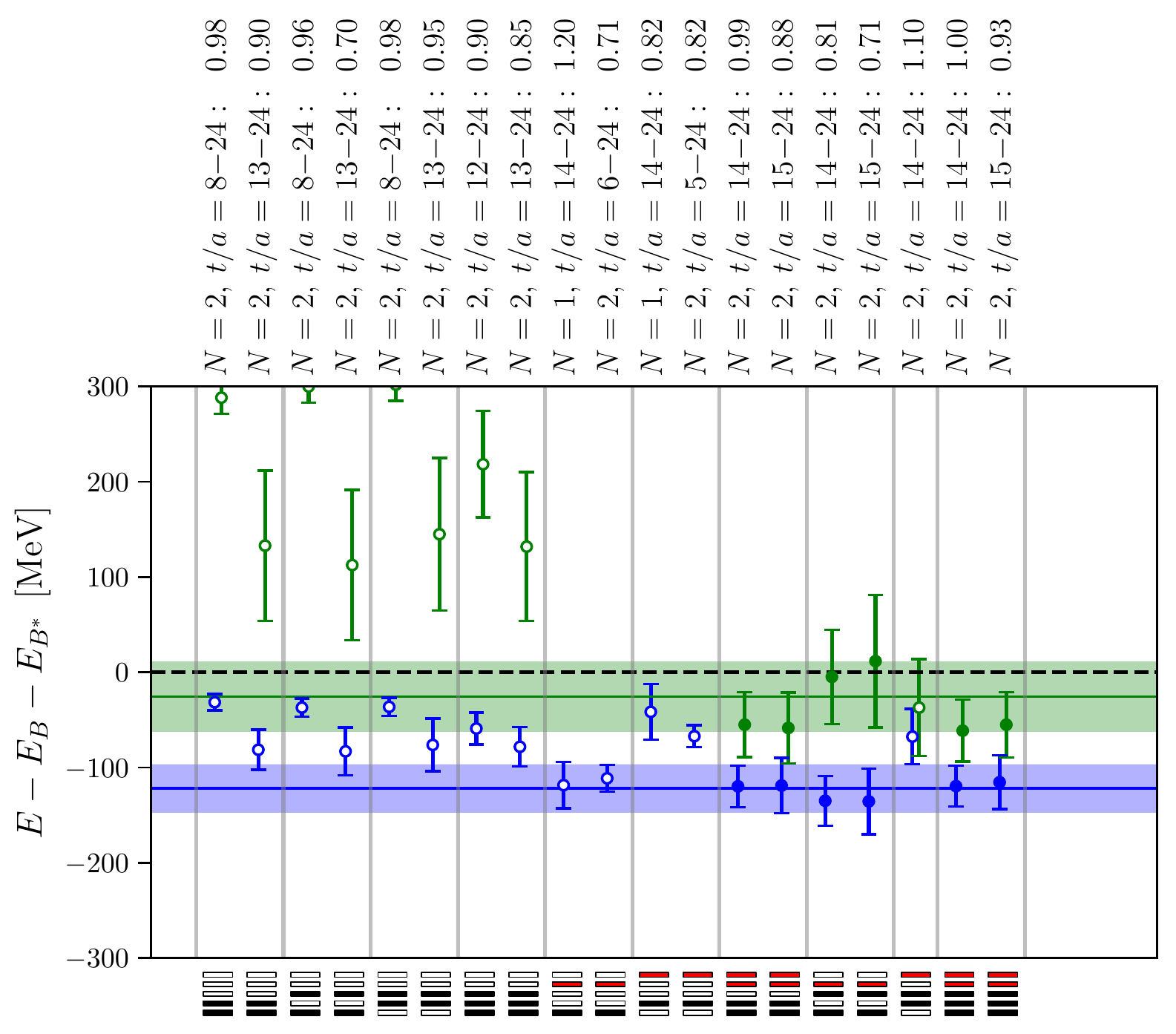}
  \caption{\label{fig:SpectrumF004} Like Fig.~\protect\ref{fig:SpectrumC005}, but for the F004 ensemble.}
\end{figure}

\begin{figure}[!htb]
  \centering
  \includegraphics[width=0.65\textwidth]{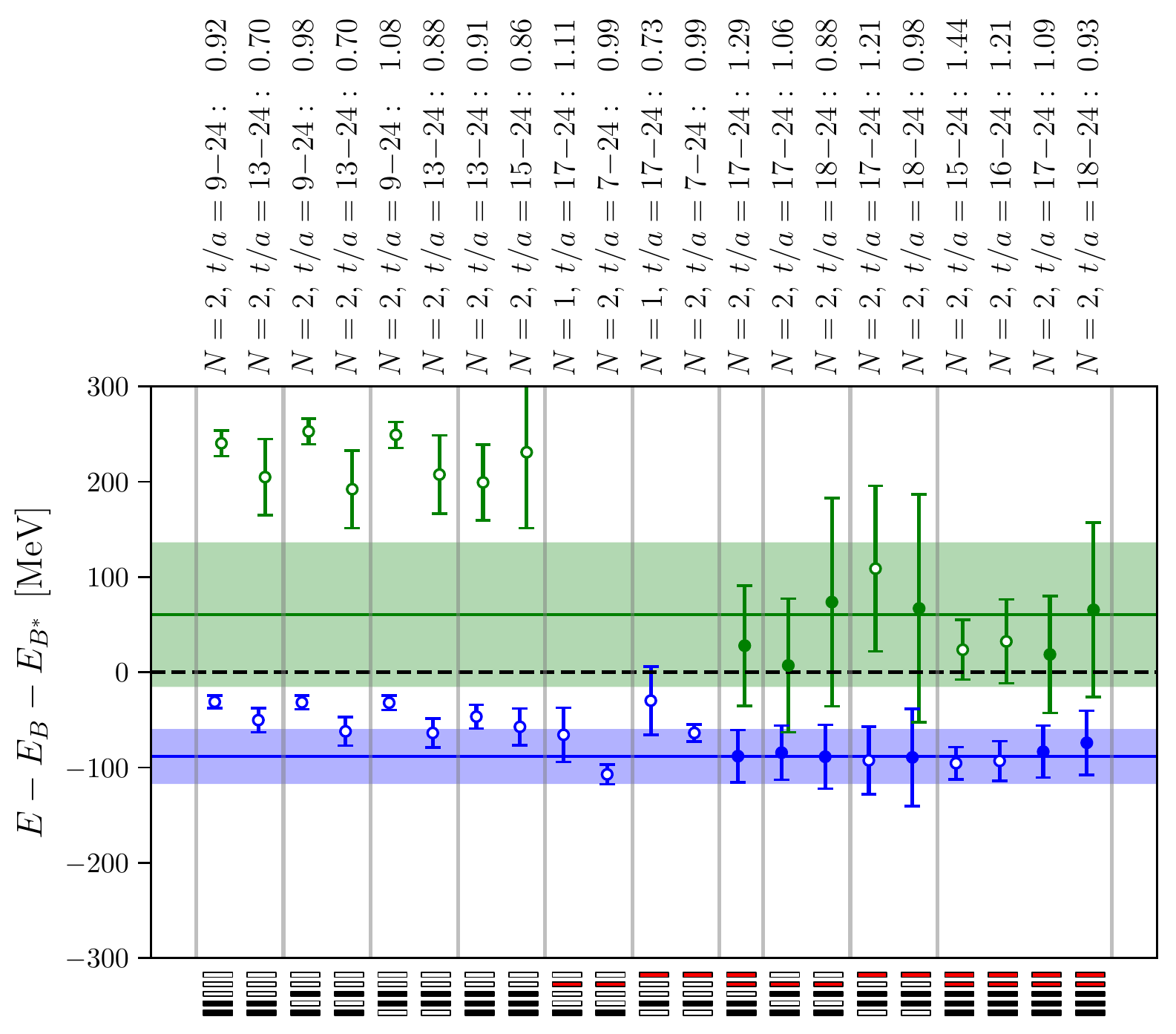}
  \caption{\label{fig:SpectrumF006} Like Fig.~\protect\ref{fig:SpectrumC005}, but for the F006 ensemble.}
\end{figure}

In Figs.~\ref{fig:SpectrumC00078}-\ref{fig:SpectrumF006} we show the two lowest energy levels for the ensembles C00078, C01, F004, and F006. The style is identical to Fig.~\ref{fig:SpectrumC005}, where the same energy levels are shown for ensemble C005, and which is discussed in detail in Sec.~\ref{sec:OverlapFactors}. The numerical results of the fits for all ensembles are given in
Tables \ref{tab:MFitresultsC00078}-\ref{tab:MFitresultsF006}.

\begin{table}
\begin{center}
\begin{tabular}{cccclll}
\hline\hline
Matrix       & \hspace{2ex} $N$ \hspace{2ex} & Fit range & \hspace{2ex} $\chi^2/{\rm d.o.f.}$ \hspace{2ex} & \multicolumn{3}{c}{$E_n-E_B-E_{B^*}$ [MeV]} \\
\hline
$(O_1, O_2)\times(O_1, O_2)$      & 2 & 7...10   & 1.05                  &  $-159(31)$, & $+61(77)$\\
$(O_1, O_2)\times(O_1, O_2)$      & 2 & 7...17   & 1.42                  &  $-136(27)$, & $+100(75)$\\
\hline
$(O_1, O_3)\times(O_1, O_3)$      & 2 & 7...10   & 0.82                  &  $-144(35)$, & $+137(85)$\\
$(O_1, O_3)\times(O_1, O_3)$      & 2 & 7...17   & 1.63                  &  $-116(30)$, & $+204(84)$\\
\hline
$(O_2, O_3)\times(O_2, O_3)$      & 2 & 7...10   & 1.32                  &  $-151(38)$, & $+104(82)$\\
$(O_2, O_3)\times(O_2, O_3)$      & 2 & 7...17   & 1.29                  &  $-106(33)$, & $+168(81)$ \\
\hline
$(O_1, O_2, O_3)\times(O_1, O_2, O_3)$       & 2 & 8...10   & 0.96                  &  $-181(40)$, & $+66(129)$     \\
$(O_1, O_2, O_3)\times(O_1, O_2, O_3)$       & 2 & 8...17   & 1.24                  &  $-125(34)$, & $+142(123)$     \\
$(O_1, O_2, O_3)\times(O_1, O_2, O_3)$       & 3 & 6...17   & 1.41                  &  $-93(24)$, & $+245(55)$, & $+612(328)$\\
$(O_1, O_2, O_3)\times(O_1, O_2, O_3)$       & 3 & 6...24   & 1.28                  &  $-86(23)$, & $+260(55)$, & $+651(337)$\\
$(O_1, O_2, O_3)\times(O_1, O_2, O_3)$       & 3 & 7...10   & 0.98                  &  $-165(31)$, & $+63(77)$, & $+6283(3782)$\\
\hline
$(O_1, O_4)\times(O_1)$      & 1 & 11...17   & 0.84                  &  $-169(96)$ \\
\hline
$(O_2, O_5)\times(O_2)$      & 1 & 11...15   & 0.84                  &  $-133(128)$ \\
\hline
$(O_1, O_2, O_4, O_5)\times(O_1, O_2)$ & 2 & 11...13   & 0.90                 &  $-181(110)$, & $-104(199)$\\
\hline
$(O_1, O_3, O_4)\times(O_1, O_3)$ & 2 & 9...15   & 1.81                 &  $-125(51)$, & $+86(98)$  \\
$(O_1, O_3, O_4)\times(O_1, O_3)$ & 2 & 9...24   & 1.52                 &  $-129(53)$, & $+79(94)$  \\
$(O_1, O_3, O_4)\times(O_1, O_3)$ & 2 & 10...16   & 1.72                 &  $-217(85)$, & $+327(187)$ \\
$(O_1, O_3, O_4)\times(O_1, O_3)$ & 3 & 5...10   & 1.29                 &  $-169(35)$, & $+13(28)$, & $+516(100)$\\
$(O_1, O_3, O_4)\times(O_1, O_3)$ & 3 & 7...10   & 1.34                 &  $-175(159)$, & $-75(143)$, & $+253(538)$\\
$(O_1, O_3, O_4)\times(O_1, O_3)$ & 3 & 5...16   & 1.55                 &  $-169(35)$, & $+7(28)$, & $+479(89)$\\
$(O_1, O_3, O_4)\times(O_1, O_3)$ & 3 & 5...24   & 1.47                 &  $-168(35)$, & $+8(27)$, & $+483(88)$\\
$(O_1, O_3, O_4)\times(O_1, O_3)$ & 3 & 7...16   & 1.55                 &  $-250(120)$, & $-76(58)$, & $+312(207)$\\
$(O_1, O_3, O_4)\times(O_1, O_3)$ & 3 & 7...24   & 1.46                 &  $-237(97)$, & $-70(52)$, & $+372(233)$\\
\hline
$(O_2, O_3, O_5)\times(O_2, O_3)$ & 2 & 8...24   & 1.30                 &  $-124(36)$, & $+0(62)$ \\
$(O_2, O_3, O_5)\times(O_2, O_3)$ & 2 & 10...24   & 1.30                 &  $-182(116)$, & $+290(173)$ \\
$(O_2, O_3, O_5)\times(O_2, O_3)$ & 3 & 5...10   & 1.52                 &  $-146(36)$, & $+5(31)$, & $+510(105)$\\
$(O_2, O_3, O_5)\times(O_2, O_3)$ & 3 & 7...10   & 1.99                 &  $-163(55)$, & $-19(111)$, & $+520(708)$\\
$(O_2, O_3, O_5)\times(O_2, O_3)$ & 3 & 5...16   & 1.40                 &  $-134(36)$, & $+2(31)$, & $+452(95)$\\
$(O_2, O_3, O_5)\times(O_2, O_3)$ & 3 & 5...24   & 1.13                 &  $-138(37)$, & $-2(30)$, & $+450(92)$\\
\hline
$(O_1, O_2, O_3, O_4, O_5)\times(O_1, O_2, O_3)$ & 2 & 10...16   & 1.34                 &  $-102(60)$, & $-54(70)$      \\
$(O_1, O_2, O_3, O_4, O_5)\times(O_1, O_2, O_3)$ & 2 & 10...24   & 1.28                 &  $-119(57)$, & $-63(71)$      \\
$(O_1, O_2, O_3, O_4, O_5)\times(O_1, O_2, O_3)$ & 3 & 6...16    & 2.61                 &  $-122(28)$, & $-21(31)$, & $+406(254)$      \\
$(O_1, O_2, O_3, O_4, O_5)\times(O_1, O_2, O_3)$ & 3 & 6...24    & 2.11                 &  $-143(54)$, & $-34(30)$, & $+359(225)$      \\
$(O_1, O_2, O_3, O_4, O_5)\times(O_1, O_2, O_3)$ & 3 & 7...16    & 1.74                 &  $-155(27)$, & $-78(39)$, & $+1736(1580)$      \\
\hline\hline 
\end{tabular}
\caption{\label{tab:MFitresultsC00078}Multi-exponential matrix fit results from the C00078 ensemble.}
\end{center}
\end{table}

\begin{table}
\begin{center}
\begin{tabular}{cccclll}
\hline\hline
Matrix       & \hspace{2ex} $N$ \hspace{2ex} & Fit range & \hspace{2ex} $\chi^2/{\rm d.o.f.}$ \hspace{2ex} & \multicolumn{3}{c}{$E_n-E_B-E_{B^*}$ [MeV]} \\
\hline
$(O_1, O_2)\times(O_1, O_2)$      & 2 & 7...22   & 0.84                  &  $-58.7(8.2)$, & $+187(15)$      \\
$(O_1, O_2)\times(O_1, O_2)$      & 2 & 14...24   & 0.87                  &  $-121(57)$, & $+535(464)$      \\
\hline
$(O_1, O_3)\times(O_1, O_3)$      & 2 & 7...22   & 0.94                  &  $-59.1(8.8)$, & $+194(15)$      \\
$(O_1, O_3)\times(O_1, O_3)$      & 2 & 14...24   & 0.75                  &  $-165(70)$, & $+571(459)$      \\
\hline
$(O_2, O_3)\times(O_2, O_3)$      & 2 & 7...22   & 1.03                  &  $-60.6(9.1)$, & $+189(15)$      \\
$(O_2, O_3)\times(O_2, O_3)$      & 2 & 14...24   & 0.99                  &  $-175(65)$, & $+602(475)$      \\
\hline
$(O_1, O_2, O_3)\times(O_1, O_2, O_3)$ & 2 & 11...24   & 0.90                 &  $-72(22)$, & $+282(79)$      \\
$(O_1, O_2, O_3)\times(O_1, O_2, O_3)$ & 2 & 14...24   & 0.80                 &  $-116(58)$, & $+628(505)$      \\
$(O_1, O_2, O_3)\times(O_1, O_2, O_3)$ & 3 & 6...24   & 0.88                 &  $-50.6(7.3)$, & $+199(11)$, & $+696(74)$      \\
$(O_1, O_2, O_3)\times(O_1, O_2, O_3)$ & 3 & 8...24   & 0.90                 &  $-57(11)$, & $+180(20)$, & $+362(264)$      \\
$(O_1, O_2, O_3)\times(O_1, O_2, O_3)$ & 3 & 14...24   & 0.78                 &  $-174(93)$, & $+550(466)$, & $+1220(1642)$      \\
\hline
$(O_1, O_4)\times(O_1)$      & 1 & 13...24   & 0.77                  &  $-74(39)$      \\
$(O_1, O_4)\times(O_1)$      & 2 & 7...24   & 0.50                  &  $-63(43)$,  &  $+113(74)$      \\
\hline
$(O_2, O_5)\times(O_2)$      & 1 & 13...24   & 0.90                  &  $-6(52)$      \\
$(O_2, O_5)\times(O_2)$      & 2 & 7...24   & 0.75                  &  $-39(34)$,  &  $+232(84)$      \\
\hline
$(O_1, O_2, O_4, O_5)\times(O_1, O_2)$ & 2 & 12...24   & 1.07                 &  $-100(26)$, & $+22(45)$      \\
$(O_1, O_2, O_4, O_5)\times(O_1, O_2)$ & 2 & 13...24   & 0.89                 &  $-94(35)$, & $-8(66)$      \\
$(O_1, O_2, O_4, O_5)\times(O_1, O_2)$ & 2 & 14...24   & 0.85                 &  $-161(49)$, & $+13(110)$      \\
\hline
$(O_1, O_3, O_4)\times(O_1, O_3)$ & 2 & 12...24   & 1.10                 &  $-92(25)$, & $+6(50)$      \\
$(O_1, O_3, O_4)\times(O_1, O_3)$ & 2 & 13...24   & 0.92                 &  $-93(35)$, & $+11(75)$      \\
$(O_1, O_3, O_4)\times(O_1, O_3)$ & 3 & 7...24   & 0.85                 &  $-59.2(8.9)$, & $+9(31)$, & $+193(16)$      \\
$(O_1, O_3, O_4)\times(O_1, O_3)$ & 3 & 11...24   & 0.79                 &  $-106(28)$, & $-38(161)$, & $+246(119)$      \\
\hline
$(O_2, O_3, O_5)\times(O_2, O_3)$ & 2 & 14...24   & 1.01                 &  $-184(57)$, & $+9(126)$      \\
$(O_2, O_3, O_5)\times(O_2, O_3)$ & 2 & 15...24   & 0.92                 &  $-164(78)$, & $+302(276)$      \\
$(O_2, O_3, O_5)\times(O_2, O_3)$ & 3 & 6...24   & 1.00                 &  $-49.9(7.8)$, & $+88(28)$, & $+204(21)$      \\
$(O_2, O_3, O_5)\times(O_2, O_3)$ & 3 & 8...24   & 0.98                 &  $-60(11)$, & $-15(50)$, & $+178(20)$      \\
$(O_2, O_3, O_5)\times(O_2, O_3)$ & 3 & 10...24   & 1.07                 &  $-65(29)$, & $-23(141)$, & $+219(47)$      \\
\hline
$(O_1, O_2, O_3, O_4, O_5)\times(O_1, O_2, O_3)$ & 2 & 11...24   & 1.25                 &  $-100(19)$, & $-8(28)$      \\
$(O_1, O_2, O_3, O_4, O_5)\times(O_1, O_2, O_3)$ & 2 & 12...24   & 1.07                 &  $-93(26)$, & $+25(45)$      \\
$(O_1, O_2, O_3, O_4, O_5)\times(O_1, O_2, O_3)$ & 2 & 13...24   & 0.96                 &  $-82(36)$, & $+9(64)$      \\
$(O_1, O_2, O_3, O_4, O_5)\times(O_1, O_2, O_3)$ & 2 & 14...24   & 0.91                 &  $-148(51)$, & $+22(110)$      \\
$(O_1, O_2, O_3, O_4, O_5)\times(O_1, O_2, O_3)$ & 3 & 10...24   & 1.47                 &  $-89(15)$, & $+40(122)$,  & $+215(67)$      \\
$(O_1, O_2, O_3, O_4, O_5)\times(O_1, O_2, O_3)$ & 3 & 11...24   & 1.08                 &  $-97(20)$, & $+136(199)$, & $+288(160)$      \\
$(O_1, O_2, O_3, O_4, O_5)\times(O_1, O_2, O_3)$ & 3 & 12...24   & 1.03                 &  $-90(27)$, & $-14(231)$, & $+360(199)$      \\
\hline\hline 
\end{tabular}
\caption{\label{tab:MFitresultsC005}Multi-exponential matrix fit results from the C005 ensemble.}
\end{center}
\end{table}

\begin{table}
\begin{center}
\begin{tabular}{cccclll}
\hline\hline
Matrix       & \hspace{2ex} $N$ \hspace{2ex} & Fit range & \hspace{2ex} $\chi^2/{\rm d.o.f.}$ \hspace{2ex} & \multicolumn{3}{c}{$E_n-E_B-E_{B^*}$ [MeV]} \\
\hline
$(O_1, O_2)\times(O_1, O_2)$      & 2 & 7...24   & 1.11                  &  $-37.6(7.2)$, & $+203(13)$      \\
$(O_1, O_2)\times(O_1, O_2)$      & 2 & 14...24   & 1.03                  &  $-104(47)$, & $-57(97)$      \\
\hline
$(O_1, O_3)\times(O_1, O_3)$      & 2 & 7...24   & 1.04                  &  $-37.7(7.5)$, & $+210(13)$      \\
$(O_1, O_3)\times(O_1, O_3)$      & 2 & 14...24   & 1.02                  &  $-108(52)$, & $-40(91)$      \\
\hline
$(O_2, O_3)\times(O_2, O_3)$      & 2 & 6...24   & 0.94                  &  $-29.0(6.8)$, & $+218(10)$      \\
$(O_2, O_3)\times(O_2, O_3)$      & 2 & 14...24   & 0.80                  &  $-84(51)$, & $-55(85)$      \\
\hline
$(O_1, O_2, O_3)\times(O_1, O_2, O_3)$ & 2 & 11...24   & 1.02                 &  $-58(17)$, & $+88(51)$      \\
$(O_1, O_2, O_3)\times(O_1, O_2, O_3)$ & 2 & 14...24   & 0.96                 &  $-104(49)$, & $-60(66)$      \\
$(O_1, O_2, O_3)\times(O_1, O_2, O_3)$ & 3 & 6...24   & 0.99                 &  $-31.1(6.5)$, & $+211(10)$, & $+825(69)$      \\
$(O_1, O_2, O_3)\times(O_1, O_2, O_3)$ & 3 & 8...24   & 1.04                 &  $-36.1(8.5)$, & $+208(18)$, & $+1281(434)$      \\
$(O_1, O_2, O_3)\times(O_1, O_2, O_3)$ & 3 & 10...24   & 1.07                 &  $-50(13)$, & $+156(36)$, & $+2821(3773)$      \\
 \hline
$(O_1, O_4)\times(O_1)$      & 1 & 13...24   & 1.16                  &  $-96(25)$      \\
$(O_1, O_4)\times(O_1)$      & 2 & 8...24   & 1.13                  &  $-114(25)$,  &  $+240(63)$      \\
\hline
$(O_2, O_5)\times(O_2)$      & 1 & 13...24   & 0.85                  &  $-70(31)$      \\
$(O_2, O_5)\times(O_2)$      & 2 & 8...24   & 1.18                  &  $-66(26)$,  &  $+394(98)$      \\
\hline
$(O_1, O_2, O_4, O_5)\times(O_1, O_2)$ & 2 & 13...24   & 1.24                 &  $-96(25)$, & $-26(41)$      \\
$(O_1, O_2, O_4, O_5)\times(O_1, O_2)$ & 2 & 14...24   & 1.10                 &  $-121(30)$, & $-90(46)$      \\
$(O_1, O_2, O_4, O_5)\times(O_1, O_2)$ & 2 & 15...24   & 1.17                 &  $-116(46)$, & $-79(84)$      \\
\hline
$(O_1, O_3, O_4)\times(O_1, O_3)$ & 2 & 13...24   & 1.17                 &  $-92(24)$, & $-12(50)$      \\
$(O_1, O_3, O_4)\times(O_1, O_3)$ & 2 & 14...24   & 0.95                 &  $-138(39)$, & $-41(66)$      \\
$(O_1, O_3, O_4)\times(O_1, O_3)$ & 3 & 7...24   & 1.19                 &  $-36.2(6.9)$, & $+30(25)$, & $+210(11)$      \\
\hline
$(O_2, O_3, O_5)\times(O_2, O_3)$ & 2 & 12...24   & 1.21                 &  $-107(22)$, & $-56(36)$      \\
$(O_2, O_3, O_5)\times(O_2, O_3)$ & 2 & 13...24   & 0.94                 &  $-79(32)$, & $-21(55)$      \\
$(O_2, O_3, O_5)\times(O_2, O_3)$ & 3 & 6...24   & 1.11                 &  $-30.0(7.5)$, & $+87(26)$, & $+244(38)$      \\
$(O_2, O_3, O_5)\times(O_2, O_3)$ & 3 & 8...24   & 1.10                 &  $-44(12)$, & $+59(32)$, & $+322(74)$      \\
$(O_2, O_3, O_5)\times(O_2, O_3)$ & 3 & 10...24   & 1.01                 &  $-123(64)$, & $-26(27)$, & $+241(116)$      \\
\hline
$(O_1, O_2, O_3, O_4, O_5)\times(O_1, O_2, O_3)$ & 2 & 13...24   & 1.14                 &  $-93(25)$, & $-21(43)$      \\
$(O_1, O_2, O_3, O_4, O_5)\times(O_1, O_2, O_3)$ & 2 & 14...24   & 1.07                 &  $-118(30)$, & $-95(47)$      \\
\hline\hline 
\end{tabular}
\caption{\label{tab:MFitresultsC01}Multi-exponential matrix fit results from the C01 ensemble.}
\end{center}
\end{table}

\begin{table}
\begin{center}
\begin{tabular}{cccclll}
\hline\hline
Matrix       & \hspace{2ex} $N$ \hspace{2ex} & Fit range & \hspace{2ex} $\chi^2/{\rm d.o.f.}$ \hspace{2ex} & \multicolumn{3}{c}{$E_n-E_B-E_{B^*}$ [MeV]} \\
\hline
$(O_1, O_2)\times(O_1, O_2)$      & 2 & 8...24   & 0.98                  &  $-31.4(8.6)$, & $+288(17)$      \\
$(O_1, O_2)\times(O_1, O_2)$      & 2 & 13...24   & 0.90                  &  $-81(21)$, & $+133(79)$      \\
\hline
$(O_1, O_3)\times(O_1, O_3)$      & 2 & 8...24   & 0.96                  &  $-37.0(9.3)$, & $+300(17)$      \\
$(O_1, O_3)\times(O_1, O_3)$      & 2 & 13...24   & 0.70                  &  $-83(25)$, & $+113(79)$      \\
\hline
$(O_2, O_3)\times(O_2, O_3)$      & 2 & 8...24   & 0.98                  & $-36.2(9.5)$, & $+302(17)$      \\
$(O_2, O_3)\times(O_2, O_3)$      & 2 & 13...24   & 0.95                  &  $-76(28)$, & $+145(80)$      \\
\hline
$(O_1, O_2, O_3)\times(O_1, O_2, O_3)$ & 2 & 12...24   & 0.90                 &  $-59(17)$, & $+218(56)$      \\
$(O_1, O_2, O_3)\times(O_1, O_2, O_3)$ & 2 & 13...24   & 0.85                 &  $-78(21)$, & $+132(78)$      \\
\hline
$(O_1, O_4)\times(O_1)$      & 1 & 14...24   & 1.20                  &  $-118(24)$      \\
$(O_1, O_4)\times(O_1)$      & 2 & 6...24   & 0.71                  &  $-111(14)$,  &  $+370(24)$      \\
\hline
$(O_2, O_5)\times(O_2)$      & 1 & 14...24   & 0.82                  &  $-42(29)$      \\
$(O_2, O_5)\times(O_2)$      & 2 & 5...24   & 0.82                  &  $-67(11)$,  &  $+495(20)$      \\
\hline
$(O_1, O_2, O_4, O_5)\times(O_1, O_2)$ & 2 & 14...22   & 0.99                 &  $-120(22)$, & $-55(34)$      \\
$(O_1, O_2, O_4, O_5)\times(O_1, O_2)$ & 2 & 15...22   & 0.88                 &  $-119(29)$, & $-58(37)$      \\
\hline
$(O_1, O_3, O_4)\times(O_1, O_3)$ & 2 & 14...24   & 0.81                 &  $-135(26)$, & $-5(49)$      \\
$(O_1, O_3, O_4)\times(O_1, O_3)$ & 2 & 15...24   & 0.71                 &  $-135(34)$, & $+12(70)$      \\
\hline
$(O_2, O_3, O_5)\times(O_2, O_3)$ & 2 & 14...24   & 1.10                 &  $-68(29)$, & $-37(51)$      \\
\hline
$(O_1, O_2, O_3, O_4, O_5)\times(O_1, O_2, O_3)$ & 2 & 14...24   & 1.00                 &  $-119(21)$, & $-61(32)$      \\
$(O_1, O_2, O_3, O_4, O_5)\times(O_1, O_2, O_3)$ & 2 & 15...24   & 0.93                 &  $-115(28)$, & $-55(34)$      \\
\hline\hline 
\end{tabular}
\caption{\label{tab:MFitresultsF004}Multi-exponential matrix fit results from the F004 ensemble.}
\end{center}
\end{table}

\begin{table}
\begin{center}
\begin{tabular}{cccclll}
\hline\hline
Matrix       & \hspace{2ex} $N$ \hspace{2ex} & Fit range & \hspace{2ex} $\chi^2/{\rm d.o.f.}$ \hspace{2ex} & \multicolumn{3}{c}{$E_n-E_B-E_{B^*}$ [MeV]} \\
\hline
$(O_1, O_2)\times(O_1, O_2)$      & 2 & 9...24   & 0.92                  &  $-31.0(6.8)$, & $+240(13)$      \\
$(O_1, O_2)\times(O_1, O_2)$      & 2 & 13...24   & 0.90                  &  $-50(13)$, & $+205(40)$      \\
\hline
$(O_1, O_3)\times(O_1, O_3)$      & 2 & 9...24   & 0.96                  &  $-31.6(7.2)$, & $+253(13)$      \\
$(O_1, O_3)\times(O_1, O_3)$      & 2 & 13...24   & 0.70                  &  $-62(15)$, & $+192(41)$      \\
\hline
$(O_2, O_3)\times(O_2, O_3)$      & 2 & 9...22   & 1.08                  & $-32.0(7.5)$, & $+249(14)$      \\
$(O_2, O_3)\times(O_2, O_3)$      & 2 & 13...24   & 0.88                  &  $-64(15)$, & $+208(41)$      \\
\hline
$(O_1, O_2, O_3)\times(O_1, O_2, O_3)$ & 2 & 13...24   & 0.91                 &  $-47(12)$, & $+199(40)$      \\
$(O_1, O_2, O_3)\times(O_1, O_2, O_3)$ & 2 & 15...24   & 0.86                 &  $-57(19)$, & $+231(80)$      \\
\hline
$(O_1, O_4)\times(O_1)$      & 1 & 17...24   & 1.11                  &  $-66(28)$      \\
$(O_1, O_4)\times(O_1)$      & 2 & 7...24   & 0.99                  &  $-107(10)$,  &  $+350(21)$      \\
\hline
$(O_2, O_5)\times(O_2)$      & 1 & 17...24   & 0.73                  &  $-30(36)$      \\
$(O_2, O_5)\times(O_2)$      & 2 & 6...24   & 0.99                  &  $-63.6(8.9)$,  &  $+479(18)$      \\
\hline
$(O_1, O_2, O_4, O_5)\times(O_1, O_2)$ & 2 & 17...24   & 1.29                 &  $-88(27)$, & $+28(63)$      \\
\hline
$(O_1, O_3, O_4)\times(O_1, O_3)$ & 2 & 17...24   & 1.06                 &  $-84(28)$, & $+7(70)$      \\
$(O_1, O_3, O_4)\times(O_1, O_3)$ & 2 & 18...24   & 0.88                 &  $-89(34)$, & $+74(109)$      \\
\hline
$(O_2, O_3, O_5)\times(O_2, O_3)$ & 2 & 17...24   & 1.21                 &  $-92(35)$, & $+109(87)$      \\
$(O_2, O_3, O_5)\times(O_2, O_3)$ & 2 & 18...24   & 0.98                 &  $-89(51)$, & $+67(120)$      \\
\hline
$(O_1, O_2, O_3, O_4, O_5)\times(O_1, O_2, O_3)$ & 2 & 15...24   & 1.44                 &  $-95(17)$, & $+24(31)$      \\
$(O_1, O_2, O_3, O_4, O_5)\times(O_1, O_2, O_3)$ & 2 & 16...24   & 1.21                 &  $-93(21)$, & $+32(44)$      \\
$(O_1, O_2, O_3, O_4, O_5)\times(O_1, O_2, O_3)$ & 2 & 17...24   & 1.09                 &  $-83(27)$, & $+19(61)$      \\
$(O_1, O_2, O_3, O_4, O_5)\times(O_1, O_2, O_3)$ & 2 & 18...24   & 0.93                 &  $-74(34)$, & $+66(92)$      \\
\hline\hline 
\end{tabular}
\caption{\label{tab:MFitresultsF006}Multi-exponential matrix fit results from the F006 ensemble.}
\end{center}
\end{table}

% ********************
% ********************
% ********************
% ********************
% ********************

\FloatBarrier
\section{\label{app:comparisonGEVPmultiexp} Comparison of multi-exponential matrix fitting and solving the GEVP}

% ********************
In this section we compare the multi-exponential matrix fitting method, used in the main part of this work, to the variational method \cite{Luscher:1990ck,Blossier:2009kd,Orginos:2015tha}. The latter is based on the generalized eigenvalue problem
\begin{align}
C_{j k}(t) v_k^n(t,t_0) = \lambda^n(t,t_0) C_{j k}(t_0) v_k^n(t,t_0),
\end{align}
where $C_{j k}(t)$ must be a hermitian square matrix. As discussed in Sec.~\ref{sec:bbudInterpolators}, our use of point-to-all propagators did not allow us to compute the correlation matrix elements $C_{j k}(t)$ with $j,k \in \{4,5\}$, which means that we can only include the operators $\op_1$ to $\op_3$ in the GEVP analysis (in contrast, the multi-exponential matrix fitting method does not require a square matrix and allows us to include $O_4$ and $O_5$ at the sink).

\begin{figure}[htb]
  \centering
  \includegraphics[width=0.45\textwidth]{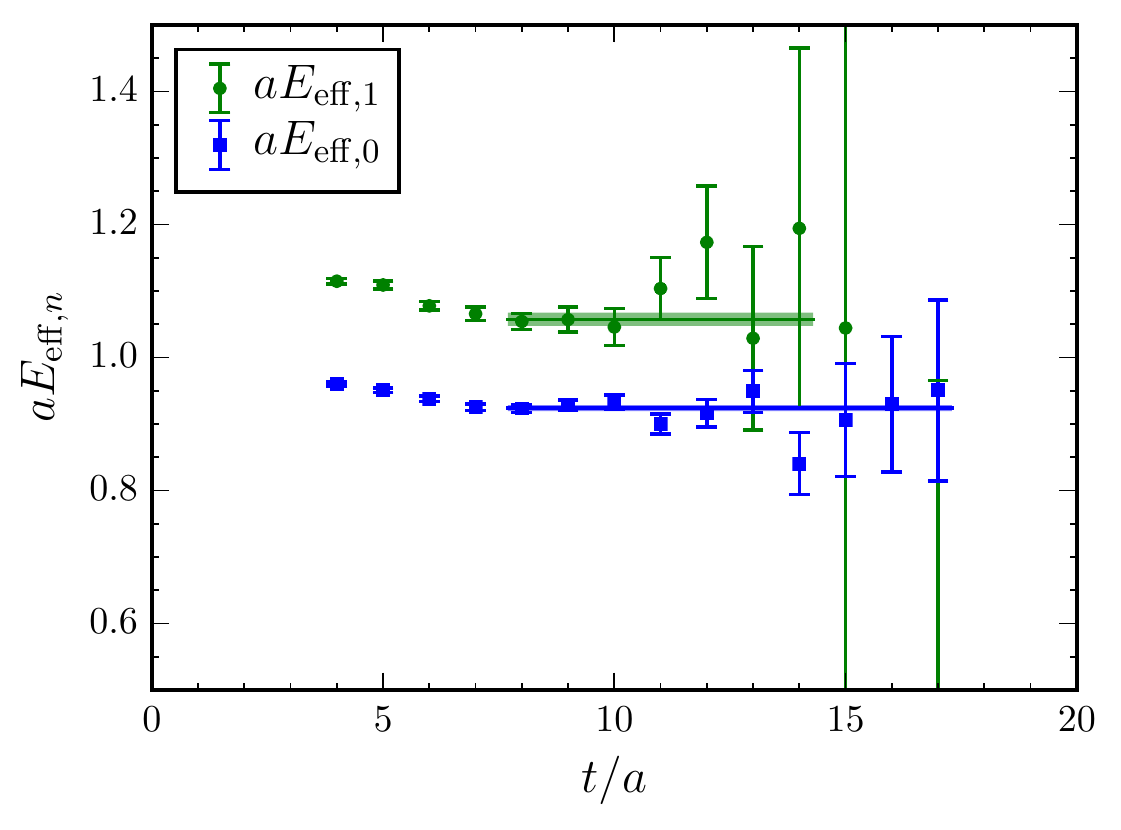}
  \caption{\label{fig:GEVPexample}Effective energies $a E_{\textrm{eff},n}$ for $n = 0, 1$ obtained for the C005 ensemble by solving the GEVP for the $3 \times 3$ correlation matrix $C_{j k}(t)$ containing operators $\op_1$ to $\op_3$. The horizontal lines show the results of constant fits in the regions $t/a = 8 \ldots 17$ ($n = 0$) and $t/a = 8 \ldots 14$ ($n = 1$).}
\end{figure}

For large time, the eigenvalues $\lambda^n(t,t_0)$ are expected to satisfy
\begin{align}
\lambda^n(t,t_0) \propto \textrm{e}^{-E_n t} ,
\end{align}
where $E_n$ is the energy of the $n$th state. We define the effective energy $E_{\textrm{eff},n}(t)$ as
\begin{align}
E_{\textrm{eff},n}(t) = \frac{1}{a} \ln\left(\frac{\lambda^n(t,t_0)}{\lambda^n(t+a,t_0)}\right),
\end{align}
which for large $t$ should plateau at $E_n$. An example for $E_{\textrm{eff},n}(t)$ is shown in Fig.~\ref{fig:GEVPexample}. We determined $E_n$ from constant fits
to $E_{\textrm{eff},n}(t)$ in a suitable range $t_\textrm{min} \leq t \leq t_\textrm{max}$ such that $\chi^2 / \textrm{d.o.f.}\lesssim 1$.
Alternatively, one can fit exponential functions $A \textrm{e}^{-E_n t}$ to the eigenvalues $\lambda^n(t,t_0)$. We performed such fits as cross-checks and found consistent energies and statistical uncertainties.

We compared the multi-exponential matrix-fit method and the GEVP method by determining the lowest two energy levels with both methods in the following way:
\begin{itemize}
\item We used the same symmetric correlation matrices: $2 \times 2$ correlation matrices with operators $(\op_1,\op_2)$, $(\op_1,\op_3)$ and $(\op_2,\op_3)$ as well as the $3 \times 3$ correlation matrix with operators $(\op_1,\op_2,\op_3)$.

\item We used the same values for $t_\textrm{min}$ (larger $t_\textrm{min}$ leads to a stronger suppression of excited states; we performed a comparison for several values for $t_\textrm{min}$).

\item We used similar values for $t_\textrm{max}$ (the results only weakly depend on $t_\textrm{max}$; we chose $t_\textrm{max}$ as the largest temporal separation where the signal is not lost in statistical noise).

\item We checked that the energy levels obtained by solving the GEVP are independent of the parameter $t_0$ for $t_0 / a = 1,2,\ldots,6$ (results shown in the following were obtained with $t_0/a = 3$).
\end{itemize}
This comparison is shown in Table \ref{tab:GEVPvsMEXP} for the C005 ensemble. It is reassuring that the results obtained from multi-exponential matrix fits and from the GEVP are in excellent agreement (when using the same operator bases). The two methods were also compared extensively in the study of $\pi\pi$ scattering in Ref.~\cite{Alexandrou:2017mpi}, where agreement was also found.

\begin{table}[htb]
\centering
  \begin{tabular}{cC{1.5cm}|C{1.6cm}C{1.2cm}l|C{1.6cm}C{1.2cm}l}
    \hline\hline
    \multirow{2}{*}{Operators} & \multirow{2}{1.5 cm}{Energy difference} &  \multicolumn{3}{c|}{Multi-exponential fitting} & \multicolumn{3}{c}{GEVP}\\ 
     & & Fit range & $\chi^2/{\rm d.o.f.}$ &  & Fit range & $\chi^2/{\rm d.o.f.}$ &  \\
    \hline
    \multirow{2}{*}{$(\op_1, \op_2)\times(\op_1, \op_2)$}      & $ \Delta E_0 $   & \multirow{2}{*}{7...22}   & \multirow{2}{*}{0.84} &  $-58.7(8.2)$
                              & 7...20            & 0.70  & $-60.5(11.4)$       \\                      
                                      & $ \Delta E_1 $  &               &       & $+187(15)$   
                              & 7...14          &  0.53 & $+183(19)$   \\
    \hline
    $(\op_1, \op_2)\times(\op_1, \op_2)$      & $ \Delta E_0 $  & 14...24           & 0.87  &  $-121(57)$ 
                              & 14...20           & 0.61  & $-129(67)$      \\                      
    \hline \hline
    \multirow{2}{*}{$(\op_1, \op_3)\times(\op_1, \op_3)$}      & $ \Delta E_0 $   & \multirow{2}{*}{7...22}   & \multirow{2}{*}{0.94} &  $-59.1(8.8)$
                              & 7...20            & 0.80  & $-62.1(12.1)$  \\                     
                                      & $ \Delta E_1 $  &               &       & $+194(15)$ 
                              & 7...14          & 0.51  & $+188(19)$   \\
    \hline
    $(\op_1, \op_3)\times(\op_1, \op_3)$      & $ \Delta E_0 $  & 14...24           & 0.75  & $-165(70)$ 
                              & 14...20           & 0.42  & $-160(76)$      \\                      
    \hline \hline
    \multirow{2}{*}{$(\op_2, \op_3)\times(\op_2, \op_3)$}      & $ \Delta E_0 $   & \multirow{2}{*}{7...22}   & \multirow{2}{*}{1.03} &  $-60.6(9.1)$
                              & 7...20            & 0.86  & $-61.4(12.3)$     \\                      
                                      & $ \Delta E_1 $  &               &       & $+189(15)$ 
                              & 7...14          & 0.56  & $+188(19)$   \\
    \hline
    $(\op_2, \op_3)\times(\op_2, \op_3)$      & $ \Delta E_0 $  & 14...24           & 0.99  & $-175(65)$ 
                              & 14...20           & 0.44  & $-164(79)$      \\                      
    \hline \hline
    \multirow{2}{*}{$(\op_1, \op_2, \op_3)\times(\op_1, \op_2, \op_3)$} & $ \Delta E_0 $ & \multirow{2}{*}{6...24}  & \multirow{2}{*}{0.88} & $-50.6(7.3)$  
                                & 6...17            & 1.32          & $-53.4(10.2)$       \\
                                           & $ \Delta E_1 $ &               &             & $+199(11)$ 
                                & 6...14          & 0.85          & $+198(14)$   \\
    \hline
    \multirow{2}{*}{$(\op_1, \op_2, \op_3)\times(\op_1, \op_2, \op_3)$} & $ \Delta E_0 $ & \multirow{2}{*}{8...24}  & \multirow{2}{*}{0.90} & $-57(11)$  
                                & 8...17            & 0.85          & $-62.8(13.7)$ \\  
                                           & $ \Delta E_1 $ &               &             & $+180(20)$ 
                                & 8...14          & 0.57          & $+176(24)$   \\
    \hline
    $(\op_1, \op_2, \op_3)\times(\op_1, \op_2, \op_3)$ & $ \Delta E_0 $ & 11...24           & 0.90    & $-72(22)$ 
                                & 11...17           & 0.73          & $-92(29)$ \\  
    \hline  
    $(\op_1, \op_2, \op_3)\times(\op_1, \op_2, \op_3)$ & $ \Delta E_0 $ & 14...24           & 0.78      & $-174(93)$  
                                & 14...17           & 0.43          & $-156(80)$ \\   
    \hline\hline    

  \end{tabular}
  \caption{\label{tab:GEVPvsMEXP}Comparison of the results for the two lowest $\bar{b}\bar{b}ud$ energy levels (relative to the $B B^\ast$ threshold, i.e., $\Delta E_n = E_n - E_B - E_{B^*}$) from multi-exponential matrix fitting and from the GEVP, for the C005 ensemble.}
\end{table}

% ********************
% ********************
% ********************
% ********************
% ********************

\FloatBarrier

\providecommand{\href}[2]{#2}\begingroup\raggedright\endgroup

% ********************

\end{document}